\documentclass[pre,floats,superscriptaddress,usenames]{revtex4}
\usepackage[T1]{fontenc}
\usepackage[latin9]{inputenc}
\usepackage{amsmath}
\usepackage{amssymb}
\usepackage{graphicx}
\usepackage{esint}

\makeatletter

\usepackage{bm} 
\usepackage{amsmath}
\usepackage{amsbsy}

\newcommand{\ttr}[1]{{\bf #1}}

\newcommand{\tr}[1]{{#1}}

\newcommand{\tg}[1]{{#1}}

\newcommand{\be}{\begin{equation}} 
\newcommand{\ee}{\end{equation}}

\newcommand{\rr}{{\bf r}}
\newcommand{\NN}{{\bf \nabla}}
\newcommand{\FF}{{\bf F}}

\newcommand{\vv}{{\bf v}}

\newcommand{\fa}{f^{\alpha}}
\newcommand{\na}{n^{\alpha}}
\newcommand{\nb}{n^{\beta}}

\newcommand{\cc}{{\bf c}}
\newcommand{\uu}{{\bf u}}
\newcommand{\uua}{{\bf u}^{\alpha}}

\newcommand{\uai}{u^{\alpha} _i}
\newcommand{\uaj}{u^{\alpha }_j}

\newcommand{\uub}{{\bf u}^{\beta}}
\newcommand{\ma} {m^{\alpha}}

\newcommand{\mb} {m^{\beta}}

\newcommand{\sab}{\sigma^{\alpha\beta}}
\newcommand{\gab}{g^{\alpha\beta}}
\newcommand{\muab}{\mu^{\alpha\beta}}

\newcommand{\bk}{\hat{\bf s}}

\begin{document}

\title{Charge transport in nanochannels: a molecular theory}

\author{Umberto Marini Bettolo Marconi}

\address{Scuola di Scienze e Tecnologie, 
Universit\`a di Camerino, Via Madonna delle Carceri, 62032,
Camerino, INFN Perugia, Italy}

\author{Simone Melchionna}
\address{CNR-IPCF, Consiglio Nazionale delle Ricerche, 
Dipartimento di Fisica, Universit\`a La Sapienza, 00185 Rome, Italy}

\begin{abstract}
We introduce a theoretical and numerical method to investigate the flow of charged fluid mixtures 
under extreme confinement. We model the electrolyte solution as
 a  ternary mixture, comprising two ionic species of opposite charge and a third uncharged component. 
The microscopic approach is based on kinetic theory and is fully self-consistent. It  allows to determine 
configurational properties, such as layering near the confining walls, and the flow properties.
We show that, under appropriate assumptions, the approach reproduces the phenomenological equations
used to describe electrokinetic phenomena, without
requiring the introduction of constitutive equations
to determine the fluxes. 
Moreover, we model channels of arbitrary shape and nanometric roughness, features that have
important repercussions on the transport properties of these systems.

Numerical simulations are obtained by solving the evolution dynamics of the one-particle
phase-space distributions of each species by means of a Lattice Boltzmann method for
flows in straight and wedged channels. 
Results are presented for the microscopic density, the velocity profiles and 
for the volumetric and charge flow-rates. Strong departures from electroneutrality
are shown to appear at molecular level.
 \end{abstract}

\maketitle


\section{Introduction}
\label{Introduction}

Understanding the motion of electrical charges in liquid solutions is at the core
of many recent advances in biological and technological applications, 
ranging from molecular separation processes, DNA sequencing,
oil recovery, decontamination of soils by  extracting and removing pollutants and toxins, 
manipulation of colloidal materials, and so forth.
Electro-nanofluidics allows controlling the movement of liquids in and around objects 
with characteristic size $H$ of the order of $10-50$ nm, 
and require a deep understanding of the transport properties
of small amounts of fluids \cite{denBerg,Sparreboom,Bocquet}.
In fact, charged liquid solutions under confinement behave in a distinctive way under the influence of electric fields, because  their flow properties depend on the charges of the moving ions and on those sitting at the pore walls.
 An ionic atmosphere oppositely charged to the surface charge and of  thickness proportional 
to the Debye length, $\lambda_D$,
 forms next to the walls, giving rise to  the so-called electric double layer (EDL).
 The difference with neutral fluids is more pronounced when the Debye screening length of the mobile 
 charges becomes comparable
with the size of the confining pores \cite{Schoch,Bruus}.

Smoluchowski provided the first explanation of mass transport of electrolytes in electro-osmotic conduits
by using the theory of the EDL of Gouy and Chapman. He computed the electric field
due to the surface charges and to the mobile ions by solving  the Navier-Stokes (NS) equation in the limit of small Reynolds
numbers. 
In electrically driven flows, the  predicted fluid velocity profile of an electrolyte displays 
a characteristic plug-like profile
departing from the  Poiseuille parabolic profile observed in pressure driven fluids \cite{Masliyah}.
The work of Nernst and Planck, instead, extended the standard theory of diffusion to include ionic systems,
by representing the ionic current  in terms of a phenomenological convection-diffusion-migration equation
\cite{Kirby}.

 Although  successful in describing microfluidic phenomena,
 the classical electrokinetic approach 
 fails to represent the behavior of nanoscopic system.
 Its major shortcoming is the 
continuum description, which  breaks down at atomistic scales, where the system behavior becomes dominated by wall effects
and it is even hard to define a bulk volume. While in microfluidics the effect of the surface charge affects
the fluid only within a distance of the order of $\lambda_D << H$,
in nanochannels the Debye length may cause the channel to become ion-selective by excluding
one type of ion over the other \cite{Mukhopadhyay}.
Therefore, some features of the continuum approach have to be modified
a) by considering the molecular nature of the solvent which competes for room in the pores and 
affects the structural properties of the EDL, and
b)  by treating the transport properties beyond the standard NS and Poisson-Nernst-Planck (PNP) 
equations which are valid in the hydrodynamic limit \cite{kontturi2008ionic}.

A variety of methods in statistical mechanics have been applied to study the transport
and structural  properties  of charged fluids from a microscopic perspective. 
These methods include Molecular Dynamics (MD)  simulations  \cite{Aluru},
Density Functional theory in conjunction with the Navier-Stokes equation \cite{Nilson}, and 
an energy variational  method \cite{Eisenberg}. 
Among these,  the nonequilibrium MD technique \cite{Singer} provides in principle  
the best description being subject to fewer approximations than other methods.
However, in the case of electrokinetic flows it has to cope with two kinds of difficulties: 
a) in the systems of interest the ion concentrations
can be orders of magnitude lower than the solvent concentration, 
so that statistical accuracy is reached only at the expenses of long computer runs,
b) the electro-osmotic speeds are much smaller than the thermal speed, 
rendering difficult to extract the signal from the background noise.

The density functional approach of Griffith and Nilson \cite{Nilson} deals rather accurately and microscopically with the configurational description of a ternary system describing an assembly
of positive and negative ions immersed in a sea of hard sphere dielectric particles mimicking water molecules, but resorts to a macroscopic
treatment of the fluid velocity introduced by means of a Navier-Stokes equation, thus neglecting the local dependence of the transport
coefficient on the density fluctuations. 
The energy variational treatment of  ions in protein channels  is similar in spirit, although derived from different premises, but
considers only a primitive model of electrolyte, that is, the solvent is not dealt explicitly  \cite{Eisenberg}. 
For complex pore geometries of physical interest
 one usually resorts to numerical tools such as finite element method, finite difference, boundary element \cite{Press}, spectral
 techniques  \cite{Karniadakis}  and recently Lattice Boltzmann methods \cite{Tian,Guo,Wangwang,Wangkang,Melchionnasucci,Melchionnaepl2011}, to mention just a few.

In the present paper we propose to treat structural and dynamical aspects of electrokinetics in confined ionic systems 
by an extension of the non-local Boltzmann-Enskog equation \cite{Vanbeijeren} 
recently developed by us for neutral multicomponent mixtures.
The method requires the knowledge of the inhomogeneous microscopic pair correlation functions of the fluid and in principle 
accommodates the same level of microscopic details as the Dynamical Density Functional theory \cite{UMBM2007,Simone2010,Rauscher2}.
The advantage  of the present approach is that both the equilibrium and the non equilibrium forces (which determine the transport 
coefficients) are derived from the same principle. 
Thus, we are able to predict not only the density profiles of the each species, but also the velocity profiles,
without resorting to further approximations. In practice, the method
brings together molecular aspects (microscopic structure, non-ideal gas effects in the equation of state, 
microscopically determined transport coefficients)  with hydrodynamic aspects of correlated fluids. 

Our theory employs an approximate method, as introduced about twenty years 
ago by Dufty and coworkers  \cite{Brey,Santos}, in order to reduce the complexity of the full collision operator
 to a form amenable to  fast iterative numerical solutions.
This  approximation is based on the idea of separating the fast degrees of freedom,
giving a minor contribution to the transport properties, from the slow hydrodynamic
modes, which on the contrary are very relevant. By this strategy one achieves 
a considerable simplification of the resulting kinetic equations
with respect to the original  Enskog-Boltzmann equations. 
This set of equations represents the interactions as self-consistent non-local forces and are amenable to
numerical solution.
In the past, numerical solutions of the Boltzmann-Enskog equation for inhomogeneous systems 
were prevented by the difficulty of solving for the distribution function in the $7$-dimensional phase-space.
This obstacle, however, can be overcome by following the same strategy 
introduced by us to handle neutral systems under inhomogeneous conditions. 
We have shown that a numerical solution method
 can be obtained in the framework of the Lattice Boltzmann approach \cite{LBgeneral,BSV92,Melchionna2008,Melchionna2009,Lausanne2010}.

The structure of this paper is the following: 
in section \ref{Model} we prepare the scene, by briefly summarizing the phenomenological 
transport equations to the mass and charge transport in pores. In section \ref{microscopic}
we set up the necessary formalism, based on an extension of the Enskog-Boltzmann transport equation,
to deal with a ternary mixture comprising positively and negatively charged hard spheres to simulate the  the ions and
uncharged hard spheres for the solvent. Next, by projecting the transport equations onto the hydrodynamic
space we  formally recover the same 
structure of equations for the charge, the mass and  the momentum as the phenomenological model.
We then obtain a microscopic expression of all the transport coefficients and include non ideal gas effects
into the theory.  We extract qualitative predictions from our theory and finally present solutions of the full
coupled system of transport equations using the LBE method.
In section \ref{Numerical},  we discuss how to solve
numerically the governing transport equations  by means of a lattice technique which bypasses the problem of
solving separately the density and momentum equations. After validating the numerical method 
using some known results, we have made some applications to show its capability to predict the conductance and 
the  volumetric flows in series of non uniform channels.
In section \ref{Conclusions} we make some conclusive remarks and considerations.

\section{Model system}
\label{Model}
The model considered here treats  the electrolyte solution
 as a ternary mixture.  Positively and negatively charged hard spheres with point like 
charges  located at their centers    represent counter- and co-ions, while
neutral spheres represents the solvent \cite{oleskyhansen}.
 Each species, denoted by $\alpha$, has mass $m^\alpha$, diameter $\sigma^{\alpha\alpha}$,
 and charge $z^\alpha e$, $e$ being the proton charge and $z^\alpha$ the ion valence.
 The index $\alpha=0$, identifies the solvent,
whose valence is zero, while
$\alpha=\pm$ identifies the two oppositely charged ionic species.
   The particles are  immersed in a medium of constant and uniform dielectric permeability, $\epsilon$,
and mutually interact via the potential
\begin{align}
U^{\alpha\beta}(r)=
   \begin{cases}
   \frac{ z^\alpha z^\beta e^2}{\epsilon r}   &\text{if }  r \geq\sigma^{\alpha\beta} \\
   \infty  &\text{if } r <\sigma^{\alpha\beta}
   \end{cases}
\end{align}

The non-coulombic part of the interaction is represented by 
an infinite repulsion which prevents the centers of any two particles to get closer than the distance
$\sigma^{\alpha\beta}=(\sigma^{\alpha\alpha}+\sigma^{\beta\beta})/2$. 
We shall not consider other types
of interactions such as attractive solvation forces between ions and solvent, which 
can be important to stabilize the solution \cite{Singer}.
  
  In the kinetic description employed here,
repulsive forces are accounted for by means of instantaneous collisions
and treated a  la Enskog, while the remaining 
forces are treated within the random phase approximation,  disregarding correlations.
Our treatment neglects the excess free energy stemming from electrostatic correlations, an approximation which can
be justified   assuming  that it is small compared to the free energy excess associated 
with the excluded volume term.
 
We first define the basic variables of our theory. Given the partial number densities $\na(\rr,t)$, 
the total number density is 
\be
n(\rr,t)=\sum_\alpha \na(\rr,t),
\ee
the  charge density $\rho_e(\rr,t)$  of the fluid is
\be
\rho_e(\rr,t)=e\sum_\alpha z^\alpha\na(\rr,t),
\ee
and the total mass density is
\be
\rho_m(\rr,t)=\sum_\alpha \ma \na(\rr,t).
\ee
We also define the average velocities of each species, $\uua(\rr,t)$,
and the average barycentric velocity as
\be
\uu(\rr,t)=\frac{\sum_{\alpha}\ma\na(\rr,t)\uua(\rr,t)}
{\rho_m(\rr,t)},
\ee
In the following, we neglect spatio-temporal variations of the temperature, $T$ for the sake of simplicity.


\subsection{Continuum equations}
\label{macroequations}

Before proceeding with the microscopic treatment, it is worth 
reviewing the macroscopic description of transport of charged liquids.
Even the simplest theoretical description of electrokinetic phenomena requires the use of three physical models:
a) the Poisson equation relating the electric field to its source, the spatial charge distribution; b) the constitutive
Nernst-Planck  equation giving  the current of each ionic species as a function of fluid velocity, charge distribution and 
applied electric field; c) the hydrodynamic continuity and Navier-Stokes  (NS) equations relating the density and fluid velocity
to the body and pressure forces acting on the system. 

The three  equations are coupled and their solution can be obtained by standard numerical methods  \cite{Masliyah}. 
Due to its low computational cost, the Poisson-Nernst-Planck equation represents the workhorse for 
diffusive electrokinetic phenomena \cite{Kirby}.
It treats the ions as point particles mutually interacting via the Coulomb potential
and experiencing a drag force stemming from the solvent. 
The solvent  does not enter explicitly the description, 
but only determines  the values of the dielectric constant
and those of the transport coefficients. 

The ionic currents, ${\bf J}^\pm \equiv n^\pm {\bf u}^\pm$, have the following phenomenological Nernst-Planck expression   
\be
{\bf J}^\pm (\rr,t)=  -D^\pm \nabla n^\pm(\rr,t)   -\lambda^\pm e z^\pm n^\pm(\rr,t) \nabla \psi(\rr,t)  +n^\pm(\rr,t) \uu(\rr,t)
\label{phenocurrent}
\ee
which is the superposition of three contributions:
 i) the one  proportional to the  gradient of the ionic concentration via the diffusion coefficient, $D^\pm$ 
 , ii)  the migration current  proportional to the gradient of the electrostatic potential  $\psi(\rr,t)$   via
the Nernst-Einstein mobility 
 $\lambda^\pm=D^\pm/(k_B T)$,  on account of the friction exerted by the solvent on the ions, 
 iii) and the convective current  due to the displacement of the fluid with velocity $\uu$. 
 In the Nernst-Planck approximation the cross coefficients that couple the diffusive transport of the different components  are neglected. In the  Stefan-Boltzmann approach, instead, such a coupling
 is preserved, but the determination of the cross-coefficients requires either a detailed microscopic theory
 or careful experimentation\cite{kontturi2008ionic}.

 The electrostatic potential  $\psi(\rr,t)$ is generated by the charge distribution $\rho_e(\rr,t)$
and by the fixed charges located on the pore surfaces and on the electrodes and satisfies the Poisson equation:
\be
\nabla^2 \psi(\rr,t)= -\frac{\rho_e(\rr,t)}{\epsilon}
\label{Poisson}
\ee
with boundary conditions $-\nabla \psi(\rr,t)=\Sigma(\rr)/\epsilon$ at the confining surfaces,
where $\Sigma(\rr)$ is the surface charge density.
The Poisson-Nernst-Planck (PNP) equation is obtained by considering the local conservation law:
\be
\frac{\partial}{\partial t} n^\pm(\rr,t)+  \NN \cdot  {\bf J}^\pm(\rr,t) = 0
\label{continuity}
\ee
with $J^\pm$ given by \ref{phenocurrent} and $\psi$ by the Poisson equation \ref{Poisson}.
To  complete the picture one determines the velocity by
considering  the  Navier-Stokes evolution  equation for  $\uu(\rr,t)$:
\be
\partial_t \uu+\uu\cdot\nabla\uu=-\frac{1}{\rho_m} \nabla P+\frac{\rho_e}{\rho_m} {\bf E}+
\frac{\eta}{\rho_m} \nabla^2 \uu +\frac{\frac{1}{3}\eta+\eta_b}{\rho_m} \nabla(\nabla\cdot \uu)
 \label{navierstokes} 
 \ee
 where $P$ is the hydrostatic pressure, $\eta$ and $\eta_b$ are the shear and bulk 
 dynamic viscosities, respectively, and ${\bf E}$ the electric field which is the sum of the
 field due to the ions and of that due to the external charges.
 
When considering a laminar flow in a straight slit-like channel, normal to the $z$ direction,  
by imposing the vanishing of the current in the $z$-direction in \ref{phenocurrent}
 one obtains the Boltzmann ionic distribution along $z$: $n^\pm(z)= c^\pm e^{-e z^\pm \psi(z)/k_B T}$,
 where $\psi(z)$ is determined with the help of  \ref{Poisson}. 
 The constants $c^\pm$ are fixed by boundary conditions, as discussed below.

The stationary solutions of   \ref{navierstokes} of great interest are when the 
electrolytes are driven through narrow channels by an external electric field.
Since the pioneering work of Smoluchowski on electro-osmotic phenomena, it is known that
when the $\lambda_D<< H$ 
the velocity is independent of the conduit size $H$, unlike pressure driven flows.
    This fact is very important and makes convenient to use electric driving to pump fluids in ultrafine capillaries.
In nanometric channels this condition is hardly realized. 
 
 For vanishing pressure gradients and under creeping flow conditions ( Stokes approximation ), 
one can use again equation \ref{Poisson} to eliminate the charge density  $\rho_e$, and obtain 
from \ref{navierstokes} the following relation between the
fluid velocity and the externally imposed electric field along the $x$ direction:
\be
\epsilon E_x\frac{\partial ^2 \psi(z)}{\partial z^2}= \eta  \frac{\partial^2 u_x(z)}{\partial z^2}
\label{vprofile0}
\ee
with $E_x=-\nabla_x \psi$.
By imposing the condition of vanishing velocity
at the surfaces of the channel, located at $z=0$ and $z=H$, one finds the celebrated Smoluchowski relation:
\be
u_x(z)=(\psi(z)-\zeta)\frac{\epsilon E_x}{\eta}
\label{vprofile}
\ee
where $\zeta$ is the value of the potential at the shear plane where the velocity vanishes \cite{Bruus}.
 
The main drawback of the macroscopic approach is that it holds only if the ions are treated as point-like particles
 migrating under   the action of the self-induced electric field, diffusing in the solvent
 and convected by the flowing solvent.
 When transport occurs through narrow channels the deviation from this ideal gas representation
 becomes crucial.
 
\section{The microscopic approach}
\label{microscopic}

In the following, we shall introduce a microscopic representation of the electrolytic solution
\tr{by switching to a kinetic description}.
We will extend the self-consistent dynamical method, that we recently employed to study 
neutral binary mixtures, to ternary charged mixtures.
We consider the following set of  Enskog-like equations   governing
the evolution of  the
\tr{one-particle phase space distributions $\fa(\rr,\vv,t)$ of each species:}
\be
\frac{\partial}{\partial t}\fa(\rr,\vv,t) +\vv\cdot\NN \fa(\rr,\vv,t)
+\frac{\FF^{\alpha}(\rr)}{\ma}\cdot
\frac{\partial}{\partial \vv} \fa(\rr,\vv,t)
= \sum_\beta\Omega^{\alpha\beta}(\rr,\vv,t) .
\label{evolution}
\ee

\tr{
The left hand side of \ref{evolution} represents the streaming contribution to the evolution, with $\FF^\alpha$
being the external force acting on species $\alpha$,
 and merely reflects
the Liouville-like single particle dynamics, while the right hand side involves the interactions among the particles
and will be the object of a series of approximations in order to render the theory numerically tractable.
The first approximation \cite{Melchionna2009} consists in
separating the interaction into an harshly repulsive term, whose effect is represented 
as collisions between hard spheres, and terms whose gradients have a slower spatial variation
and can be treated within a mean-field like approximation.}
\tr{According to the Enskog approach the effect of repulsion is obtained by  considering pairs of particles 
dynamically uncorrelated
before collisions, but with probabilities modulated by the static pair  distribution function.
The velocities of these pairs, however,  emerge correlated  after a collision. 
The collision process is described within the 
 revised Enskog theory (RET) proposed by van Beijeren
and Ernst \cite{Vanbeijeren} which is known to give good results for the transport properties of hard spheres.}
\tr{Since even the RET operator $\Omega^{\alpha\beta}$ (see Ref.\cite{Lausanne2010} for an explicit representation) is awkward to manage,
 simplifying methods are very convenient:
the RET  operator  is further simplified by  invoking the approximation put forward
by Dufty and coworkers \cite{Brey,Santos} which 
separates the hydrodynamic from the non-hydrodynamic contributions to $\Omega^{\alpha\beta}$ 
and employs a projection technique.
Extending such a prescription to the present case we obtain:} 
\begin{align}
&
\sum_\beta\Omega^{\alpha\beta}(\rr,\vv,t)\approx \nonumber\\
 & -\omega[ \fa(\rr,\vv,t)- \phi^{\alpha}_{\perp}(\rr,\vv,t)]+\frac{{\bm\Phi}^{\alpha}(\rr,t) }{k_B T} \cdot(\vv-\uu(\rr,t)) 
 \phi^{\alpha}(\rr,\vv,t) -\frac{e z^\alpha }{\ma}\NN \psi(\rr)\cdot
\frac{\partial}{\partial \vv} \fa(\rr,\vv,t)
\label{evolution2}
\end{align}

\tr{
The term ${\bm \Phi}^{\alpha}$ represents the  sum of the internal  forces of non electrostatic nature acting
 on  species $\alpha$. In addition, the electric potential $\psi$ is given by the solution
of \ref{Poisson}. 
Unlike the full collision operator of Enskog type,
the last two terms in the right hand side of Eq. \ref{evolution2} act only
 on the hydrodynamic components of the distribution function $\fa$. The non-hydrodynamic modes
 instead are driven by the first term in the right hand side of \ref{evolution2}, which is a   
Bhatnagar-Gross-Krook (BGK) type of relaxation kernel.
Its role is to drive the system towards  local equilibrium over a time scale $\omega^{-1}$,
much shorter than the hydrodynamic time scales\cite{BGK}, while the global equilibrium 
is controlled by the collisional force mentioned above.
}

The BGK term contains the
distributions functions $\phi^{\alpha}_{\perp}$ and $\phi^{\alpha}$
which have the following representations (see ref. \cite{JCP2011} for details):
\be
\phi^{\alpha}(\rr,\vv,t)=\na(\rr,t)[\frac{\ma}{2\pi k_B T}]^{3/2}\exp
\Bigl(-\frac{\ma(\vv-\uu(\rr,t))^2}{2 k_B T} \Bigl)
\label{psia}
\ee
and
\be
\phi^{\alpha}_{\perp}(\rr,\vv,t)=\phi^{\alpha}(\rr,\vv,t) \Bigl\{1+
\frac{\ma(\uua(\rr,t)-\uu(\rr,t))\cdot(\vv-\uu(\rr,t))}{k_B T}\Bigl\}
\label{prefactor}
\ee
\tr{Notice that in the case of a one-component fluid there is no difference between $\phi^{\alpha}_{\perp}$ and $\phi^{\alpha}$,
since the velocities $\uua$ and $\uu$ coincide and the standard  BGK approximation
involves the difference between the distribution $\fa$ and the
local  Maxwellian $\phi^\alpha$. 
The above prescription fullfils the indifferentiability principle which states that
when all physical properties of the species are identical, the total distribution
$f=\sum_\alpha \fa$ must obey the single species transport equation.
}

\tr{The reason to use the modified distributions   \ref{prefactor} instead of \ref{psia}  
is to obtain the correct mutual diffusion and hydrodynamic properties starting from
\ref{evolution}. A simple BGK recipe, that is, by setting $\phi^{\alpha}_{\perp}=\phi^{\alpha}$,
would lead to  double counting  the interactions on the diffusive properties.}

Equations \ref{evolution,evolution2} can be solved after specifying the forces
$\FF^\alpha$  and ${\bm\Phi}^{\alpha}$, the Coulombic force $-e\NN \psi(\rr)$
and the fluid velocities $\uu$ and $\uua$.

From the
knowledge of the $\fa$'s it is possible to determine not only all the hydrodynamic fields of interest, but
also the structure of the fluid at molecular scale. 
All equations which constitute the building blocks of the classical 
electrokinetic approach can be derived from the set of equations \ref{evolution,evolution2}.
In fact, the PNP and NS equations can be straightforwardly recovered 
by taking the appropriate velocity moments of the kinetic Enskog-Boltzmann equation. 
From the distributions, we compute the partial densities $\na(\rr,t)$,
\be
\na(\rr,t)=\int d\vv \fa(\rr,t)
\ee
the average velocities $\uua(\rr,t)$  of the species $\alpha$,
\be
\na(\rr,t)\uua(\rr,t)=\int d\vv \vv\fa(\rr,t),
\ee
and the kinetic contribution of component $\alpha$ to the pressure tensor,
\be
\pi_{ij}^{\alpha}(\rr,t)=\ma\int d\vv (v_i-u_i)(v_j-u_j)\fa(\rr,\vv,t).
\label{pressurekin}
\ee
\tr{The explicit form of the internal forces ${\bm \Phi}^{\alpha}(\rr,t)$  featuring in \ref{evolution2} is determined 
by taking the first moment of the collision integral with respect to the velocity.  By algebraic manipulations one observes that }
 ${\bm \Phi}^{\alpha}(\rr,t)$ is due to two-particle correlations
and can be separated into three contributions \cite{Melchionna2009}, as
 \be
{\bm \Phi}^{\alpha}(\rr,t)= \FF^{\alpha,mf}(\rr,t)+\FF^{\alpha,drag}(\rr,t)+\FF^{\alpha,visc}(\rr,t) .
\label{splitforce}
\ee
The first term arises from the gradient of the hard sphere chemical potential excess over the ideal gas
value:
\be
\FF^{\alpha,mf}(\rr,t)=-\nabla\mu^\alpha_{exc}(\rr,t)
\label{ventuno}
\ee
with
\be
\nabla\mu^\alpha_{exc}
=k_B T\sum_\beta(\sab)^2 
\int d\bk \bk
g_{\alpha\beta}(\rr,\rr+\sab\bk,t)
n_{\beta}(\rr+\sab\bk,t)
\label{potmeanforce}
\ee
where the integration is over surface of the unit sphere and $\gab(\rr,\rr+\sab\bk,t)$ is the inhomogeneous pair correlation functions 
for  hard spheres at contact ($|\rr-\rr'|=\sab$).
The second term accounts for the frictional force between the ions
and the solvent and depends linearly on their relative velocities:
\be
\FF^{\alpha,drag}(\rr,t)= 
-\sum_\beta {\bm \gamma}^{\alpha\beta} (\rr,t)  (\uua(\rr,t)-\uub(\rr,t))
\label{dragforce}
\ee
via the inhomogeneous friction tensor:
\be
\gamma_{ij}^{\alpha\beta}(\rr,t)=2(\sab)^2 \sqrt{\frac{2\muab k_B T}{\pi} }
\int d\bk s_i s_j
\gab(\rr,\rr+\sab\bk,t)
\nb(\rr+\sab\bk,t),
\label{tensorgamma}
\ee
where  $\muab$ 
is the reduced mass $\muab=\frac{\ma \mb}{\ma+\mb}$ for the colliding pair.
\tr{Notice the local character of the relation between the drag force and the fluid velocity.}
Finally, the particles experience a viscous force which has the following Enskog form:
\be
\FF^{\alpha,visc}(\rr,t)=
\sum_\beta 2(\sab)^2 \sqrt{\frac{2\muab k_B T}{\pi} }
\int d\bk \bk
g_{\alpha\beta}(\rr,\rr+\sab\bk,t)
\nb(\rr+\sab\bk,t)
 \bk\cdot
(\uub(\rr+\sab\bk)-\uub(\rr)) .
\label{viscousforce}
\ee
Since there is no exact theory of the pair correlation function in spatially inhomogeneous systems,
the contact value of $\gab$  is computed using 
a two-component generalization of the Fischer and Methfessel prescription\cite{Fischer}:
one approximates  the inhomogeneous  $\gab$ by the corresponding bulk  value
evaluated when the partial densities take on the values of the smeared densities,
$\bar \na(\rr)$:
\be
\gab(\rr_\alpha,\rr_\beta;|\rr^{\alpha}-\rr^{\beta}|=\sab)=
\gab_{bulk}( \{\bar \xi_n({\bf R}^{\alpha\beta})\}),
\ee
where the functions 
$$
\bar \xi_n({\bf R}^{\alpha\beta}) =\frac{\pi}{6}\sum_{\alpha}\bar n^\alpha
({\bf R}^{\alpha\beta}) (\sigma^{\alpha\alpha})^n
$$
are linear combinations of the smeared densities
$$
\bar \na({\bf R}^{\alpha\beta})=\frac{1}{V^\alpha}\int_{V^\alpha}d\rr 
\na(\rr-{\bf R}^{\alpha\beta})
$$
over spheres of volume
$V^{\alpha}=\pi (\sigma^{\alpha\alpha})^3/6$
 centered at the point  $
{\bf R}^{\alpha\beta}=
\frac{\rr^\alpha+\rr^\beta }{2}
$, where $\rr^\alpha$ and  $\rr^\beta$ are the centers of the two spheres.
The explicit expression of $\gab_{bulk}$ is provided by
an extension of the  Carnahan-Starling equation to mixtures \cite{Boublik,Mansoori}:

\be
\gab_{bulk}(\{ \xi_n\})
=\frac{1}{1-\xi_3}+\frac{3}{2}
\frac{\sigma^{\alpha\alpha}\sigma^{\beta\beta}}{\sab}\frac{\xi_2}{(1-\xi_3)^2}
+\frac{1}{2} \Bigl(\frac{\sigma^{\alpha\alpha}\sigma^{\beta\beta}}{\sab}\Bigl)^2
\frac{\xi_2^2}{(1-\xi_3)^3}.
\label{carnahan}
\ee

We remark that  the present method requires
the knowledge of the pair correlation function at contact, in contrast with the DDFT approach
which requires instead the knowledge of free energy or equivalently the Ornstein-Zernike direct correlation function. 
\tr{However, in the spirit the method here employed is equivalent to the so called Weighted Density Approximation
of equilibrium DFT.}
The latter does not contain information necessary to compute  the transport coefficients,
that in the present method are obtained from $\gab$ via the Enskog
ansatz for the collision integrals.


\subsection{Derivation of the equations of electrokinetics from the microscopic approach}

It is useful to show that in the limit of slowly varying fields 
the transport equations \ref{evolution} reproduce
the macroscopic evolution discussed in section \ref{macroequations}.
To this purpose, we integrate \ref{evolution} w.r.t. the velocity and obtain the conservation law for the
particle number of each species:
\be
\frac{\partial}{\partial t}\na(\rr,t) =-
\nabla\cdot \Bigl(\na(\rr,t)(\uua(\rr,t)- \uu(\rr,t)\Bigl)-\nabla\cdot \Bigl(\na(\rr,t) \uu(\rr,t)\Bigl)= -\nabla \cdot {\bf J}^\alpha(\rr,t)  \, ,
\label{continuityb}
\ee
In order to recover the PNP equation we consider the 
the momentum balance equation for the species $\alpha$, which is obtained 
after multiplying by $\vv$ and
integrating w.r.t. $\vv$:
\begin{align}
&
\frac{\partial}{\partial t}[\na(\rr,t)\uaj(\rr,t)]+ 
\nabla_i \Bigl(\na(\rr,t) \uai(\rr,t) \uaj(\rr,t)
-  \na(\rr,t)(u^{\alpha}_i(\rr,t)-u_i(\rr,t))( u^{\alpha}_j(\rr,t)-u_j(\rr,t))\Bigl)=
\nonumber\\
&
-\nabla_i \frac{ \pi_{ij}^{\alpha}(\rr,t)}{\ma}+ \frac{F^{\alpha}_j(\rr)}
{\ma}\na(\rr,t)+
 \frac{ \Phi^{\alpha}_{j}(\rr,t)}{\ma}\na(\rr,t) -\frac{e z^\alpha}{\ma}\na(\rr,t)\nabla_j \psi(\rr,t)
\label{momentcomponent}
\end{align}
Using \ref{ventuno}
we define the  gradient  of the global chemical potential of the individual species, $\mu^{\alpha}$,
as the sum of the ideal gas part and the excess part:
\be
n^\alpha(\rr,t)\nabla_i \mu^{\alpha}(\rr,t) =\nabla_j \pi_{ij}^{\alpha}(\rr,t) \delta_{ij} -n^\alpha(\rr,t) F_i^{\alpha,mf}(\rr,t) \, .
\ee
By assuming the existence of a  steady current, we drop the non-linear terms in the velocities in 
the l.h.s. of \ref{momentcomponent} and neglect the viscous  force contribution.
The following approximated force balance is obtained
\be
\nabla \mu^{\pm}(\rr,t) -\FF^{\pm}(\rr) +e z^\pm \nabla  \psi(\rr,t)
\approx \FF^{\pm,drag} (\rr,t) .
\label{momentcomponent2}
\ee
The r.h.s. of  \ref{momentcomponent2} represents the  drag force exerted on the particles of type $\alpha=\pm$,
in reason of their different drift velocities. In dilute solutions  the
charged components are expected to experience a large friction arising only from the solvent 
while a negligible friction  from the oppositely charged species, so that we further approximate
\be
\FF^{\pm,drag} (\rr,t) = -\gamma^{\pm} (\rr,t)  (\uu^\pm(\rr,t)-\uu(\rr,t))
\label{dragforce}
\ee
where we set $\uu\approx \uu^0$. 
In \ref{dragforce}, the friction can be evaluated in uniform bulk conditions to be
\be
\gamma^{\pm} \approx \frac{8}{3} \sqrt{2 \pi  k_B T  \frac{m^{\pm} m^0}{m^{\pm} + m^0 }}
g^{0\pm}n^0 (\sigma^{0\pm})^2 
\ee
with $n^0$ being the bulk density of the solvent and 
$g_{0\pm}$ the bulk ion-solvent pair correlation function evaluated at contact.
Finally, using  eqs. \ref{continuityb} and \ref{dragforce}  we obtain
an  expression for the  ionic currents in terms of the microscopic parameters:
 \be
{\bf J}^\pm_i(\rr,t)=  -\frac{1}{\gamma^\pm} n^\pm(\rr,t) \nabla \mu^{\pm}(\rr,t) 
-\frac{1}{\gamma^\pm}   e z^\pm n^\pm(\rr,t)\nabla \psi(\rr,t)  + n^\pm(\rr,t) \uu(\rr,t)
 \label{microcurrent}
\ee
which has the same form as the phenomenological Planck-Nernst current \ref{phenocurrent}, with the full chemical potential gradient 
$\mu^\pm$ replacing 
the ideal gas chemical potential gradient, $k_B T \nabla \ln(\na)$ and $\gamma^\pm=k_B T/D^\pm$
and $\lambda^\pm=1/\gamma^\pm$.
 The total electric charge density current is
\be
{\bf J_e}=
 -\sum_\pm \frac{e z^\pm}{\gamma^\pm}  n^\pm(\rr,t)\nabla \mu^\pm(\rr,t) +  \sigma_{el} {\bf E}  +\rho_e(\rr,t) \uu(\rr,t)
\ee
where the zero frequency electric conductivity $\sigma_{el}$ is given by the Drude-Lorentz-like formula:
\be
\sigma_{el}=e^2  \Bigl( \frac{(z^+)^2}{\gamma^+ }n^+(\rr,t)+   \frac{(z^-)^2}{\gamma^- } n^-(\rr,t) \Bigl)
\label{Drude}
\ee
showing that the conductivity is due to collisions with the solvent and decreases as the solvent becomes denser
($\gamma^\pm$ is an increasing function of $n^0$) while increases with the number of charge carriers.
\tr{It should be noticed that, in the approximation of a local friction, 
the Drude-Lorentz conductivity is local in space and time. 
However, in the numerical solution this local approximation is not imposed}.


To derive the total momentum equation, already introduced on a phenomenological basis 
in  \ref{navierstokes}, we sum \ref{momentcomponent} over the three components:
\begin{align}
&\partial_{t}u_j(\rr,t)+ u_i(\rr,t)\nabla_i u_j(\rr,t)+\frac{1}{\rho_m}\nabla_i \pi^{(K)}_{ij}
+\frac{1}{\rho_m}\sum_\pm e z^\pm n^\pm(\rr,t)\nabla_j \psi(\rr,t)
\nonumber\\
& -\frac{1}{\rho_m} \sum_{\alpha=0,\pm} n^\alpha(\rr,t) \Bigl(F^{\alpha }_j(\rr) +F^{\alpha,mf}_{j}(\rr,t) +F^{\alpha,visc}_{j}(\rr,t)  \Bigl)=0.
\label{globalmomentumcont}
\end{align}
where
$
\pi_{ij}^{(K)}(\rr,t)=
\sum_\alpha \pi_{ij}^{\alpha}(\rr,t)
$ is the total kinetic pressure which can be approximated as:
\be
\nabla_i \pi^{(K)}_{ij} \simeq \delta_{ij}\nabla_j P_{id}
 -\eta^{(K)}\Bigl(\frac{1}{3}\nabla_i\nabla_j u_i +\nabla_i^2 u_j\Bigl) ,
\ee
with $P_{id}=k_B T\sum_\alpha \na$
and $\eta^{(K)}=\frac{k_B T}{\omega}\sum_\alpha\na$.
Using the result of ref. \cite{Melchionna2009}, we can write
\be
\sum_\alpha\na(\rr,t)\FF^{\alpha,visc}(\rr,t)    
\simeq-\eta^{(C)}\nabla^2\uu-(\frac{1}{3}\eta^{(C)}+\eta_b^{(C)})\nabla(\nabla\cdot\uu).
\ee
The non-ideal contribution to the shear viscosity is  evaluated in uniform bulk conditions to be
\be
\eta^{(C)}=\frac{4}{15}\sum_{\alpha\beta}\sqrt{2\pi\muab k_B T}(\sab)^4 \gab \na_0\nb_0 ,
\ee
while the bulk viscosity is
$\eta_b^{(C)}=\frac{5}{3}\eta^{(C)} $.
In conclusion,  \ref{globalmomentumcont} can be cast in the Navier-Stokes form
\be
\partial_{t}u_j+ u_i\nabla_i u_j=-\frac{1}{\rho_m}\nabla_i P \delta_{ij}
-\frac{\rho_e}{\rho_m}\nabla_j \psi +\frac{\eta}{\rho_m} \nabla_i\nabla_i u_j
+\frac{\frac{1}{3}\eta+\eta_b}{\rho_m} \nabla_j \nabla_i u_i
\label{globalmomentumcont2}
\ee
with $\eta=\eta^{(K)}+\eta^{(C)}$ and $\nabla_i P=\nabla_i P_{id}+\sum_\alpha n^\alpha(\rr,t)\nabla_i \mu^{\alpha}_{exc}(\rr,t)$.

Having established the correspondence between the macroscopic equations and those derived from 
the kinetic approach, we can now outline the method of solution. 


\subsection{Entropic contribution to the Poisson-Boltzmann equation in the non primitive model}

 
 Some recent studies on the non primitive model of ionic solution lead
 to a  modified Poisson-Boltzmann equation \cite{Joly,Kilic}.
 In the same spirit we derive within the present approach 
 a similar equation. We consider only mixtures of equal sizes.
At equilibrium, 
\tr{that is, when all the currents vanish,}
 according to  \ref{microcurrent}  the ionic densities satisfy the equations:
\be
k_B T  \nabla\ln n^\pm(\rr,t)+\nabla\mu_{exc}(\rr,t) +e z^\pm\nabla \psi(\rr)=0
\ee
while the solvent  satisfies the equation:
\be
k_B T  \nabla\ln n^0(\rr,t)+\nabla\mu_{exc}(\rr,t) =0
\ee
Notice  that since for equi-sized molecules $\mu^{exc}$ depends only on the overall local packing fraction
one can write:
\be
k_B T \Bigl( \nabla\ln(n^\pm(\rr,t)-\nabla\ln(n^0(\rr,t)\Bigl) +e z^\pm\nabla \psi(\rr)=0
\ee
and by integrating one gets an expression relating the ionic densities  to the solvent density
\be
n^\pm(\rr,t)=C^\pm n^0(\rr,t)\exp(-\frac{e z^\pm \psi(\rr)}{k_B T})
\label{jolyrelation}
\ee
where $C^\pm$ is an integration constant to be found by normalization.

As a result the electrostatic potential is coupled to the solvent density via the Poisson equation:
\be
\nabla^2 \psi(\rr,t)=   -\frac{e}{\epsilon}n^0(\rr,t)\sum_\pm z^\pm C^\pm  \exp(- \frac{e \sum_\pm z^\pm \psi(\rr,t)}{k_B T})
\ee
In the case of the restricted primitive model (no explicit solvent) the density of the "ghost" solvent is reabsorbed 
in the definition of the integration constant.
The numerical test of such a relation is given below in section \ref{Results}.


\section{Numerical method}
\label{Numerical}
The numerical solution of the coupled  equations for the distribution functions $\fa$, eqs. \ref{evolution},
is achieved in the framework of the Lattice Boltzmann method, as presented 
in refs. \cite{Melchionna2009,JCP2011}. 
The numerical method is a substantial modification of the conventional method used in fluid dynamics 
applications to the presence of hard sphere collisions \cite{LBgeneral}.
In a nutshell, the Lattice Boltzmann method is based on the following steps.
One discretizes the position coordinate, $\rr$, by introducing a Cartesian mesh whose lattice points 
are separated by a distance $a$. The continuous velocity, $\vv$, is also discretized by restricting 
its values to $Q$ possible "states", $\vv \to \{\cc_p\}$ with $p=1,N$. 
The discrete velocities are chosen as to connect neighboring spatial mesh points.
For the present three-dimensional study, we use a 19-speed lattice, consisting of one speed
$\cc_p= 0$ (particle at rest on a mesh node), six discrete velocities with $|\cc_p|=1$ and pointing towards
the first mesh neighbors, and $12$ particles with $|\cc_p|=\sqrt 2$ pointing towards the second mesh neighbors.
The  continuous phase space distribution functions $\fa$ 
are replaced by the array $\fa(\rr,\vv,t)\to \fa_p(\rr,t)$  and
the velocity moments are evaluated as \cite{LBgeneral}
\be
\na(\rr,t)=\sum_p \fa_p
\ee
and
\be
\na(\rr,t) \uu^\alpha(\rr,t)=\sum_p \cc_p\fa_p(\rr,t)
\ee

The modification of the conventional Lattice Boltzmann method to the case of hard sphere dynamics
is based on computing the forces encoded by eqs. \ref{potmeanforce}, \ref{viscousforce} and \ref{dragforce} 
via numerical quadratures. 
The presence of Coulomb forces requires the solution of the
Poisson equation for the electrostatic potential generated by the  mobile and surface charges. 
Its determination is achieved by using a successive over-relaxation method \cite{Press}. 
The speed of convergence of the Poisson solver is greatly enhanced by employing a 
Gauss-Siedel checker-board scheme in conjunction with Chebychev acceleration \cite{Press}. 
Neumann boundary conditions on the gradient of the electrostatic potential are imposed at the wall surface
\be
\hat n \cdot \nabla \psi |_{\rr \in S}=-\frac{\Sigma(\rr)}{\epsilon}
\ee
where $\hat n$ is the normal to the surface, $S$. 
We impose an external voltage by emulating the presence of electrodes in real systems.
The method implies imposing the value of the external electric field $E_x$ as the boundary condition 
\be
\nabla \psi=-E_x
\ee
at locations $x=0^+$ and $x=L^-$.

The electrostatic boundary conditions are imposed by using a second-order finite difference scheme, 
compatible with the overall accuracy of the Lattice Boltzmann method. 
The system is further taken to be periodic along the $x$ direction for the fluid populations of each component.
This choice implies that all hydrodynamic fields (density, currents, etc.) are periodic along the x direction.
In order to accommodate the presence of the electrodes, the hydrodynamic fields exhibit a finite 
jump at the $x=0,L,2L,...$ locations. Preliminary calculations have shown that 
this jump does not have major impact on the stationary quantities, in particular as regarding those in the pore 
region.
No slip boundary conditions on the fluid velocities of each species are imposed by means of the bounce-back 
rule \cite{LBgeneral}.
Once all forces on the right hand side of  \ref{evolution} are computed, we evolve the distribution 
function by a second-order accurate trapezoidal integration \cite{PRE2012}.
In addition, we compensate the harsh internal forces by auxiliary fields properly chosen. 
This method improves the accuracy and robustness of the numerical scheme without altering
the structure and dynamics of the system. Full details can be found in ref. \cite{PRE2012}

\section{Results}
\label{Results}

\begin{figure}[htb]
\includegraphics[clip=true,width=9.0cm, keepaspectratio]{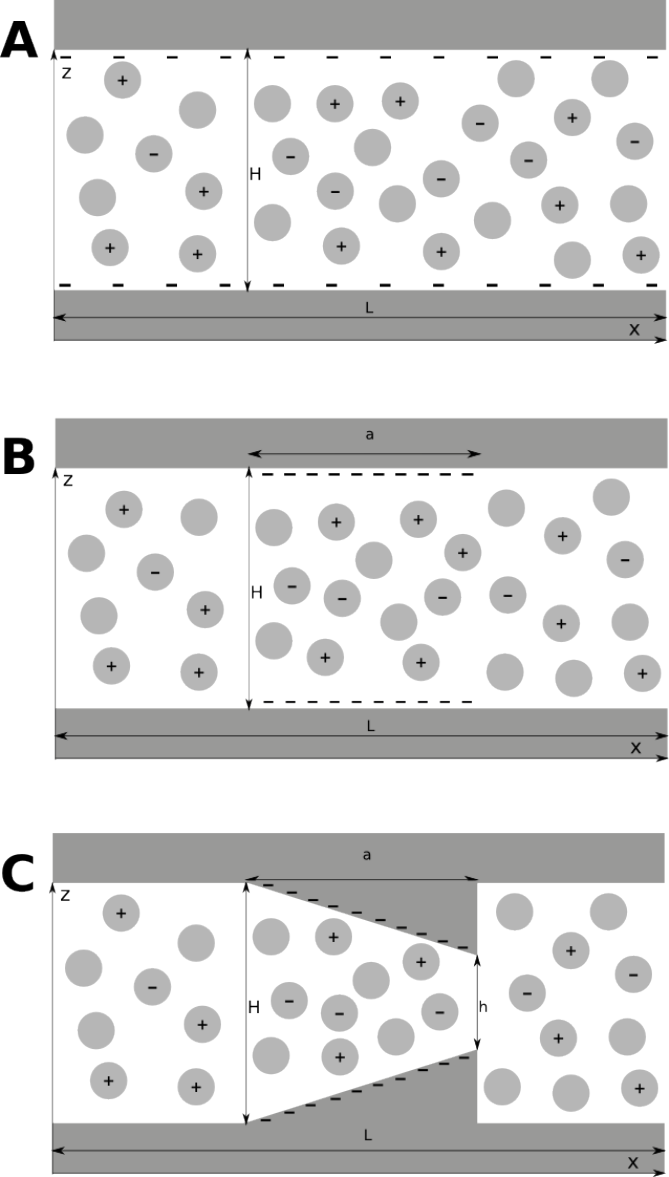}
\caption{Geometries of the model systems A, B and C, with the shaded regions representing the wall. 
Each system contains three species of hard spheres of equal diameters $\sigma$ and masses $m$, 
but different charges: $0, 1, -1$, respectively. \tg{In model A the ternary fluid moves in a channel whose walls are uniformly and negatively charged.
The system is globally neutral: $N^{+}-N^{-}=-2  \Sigma w L/e$, where $N^+$ and $N^-$ are the number of positive and negative mobile charges, respectively, $\Sigma$ is the surface charge density, and $w$ the {transversal} dimension of the channel and $L$ its length. 
Model B mimics a parallel slit channel, whose walls are decorated with negative surface charge density $\Sigma$ in a region of length  $a$.
Charge neutrality ,  $N^{+}-N^{-}=-2  \Sigma w\, a/e$, holds. In Model C a wedge shaped region of length $a$ connects two parallel slits.
The surface charge sits only on the inclined walls and $N^{+}-N^{-}=-2  \Sigma w\, \sqrt{a^2+(H-h)^2/4}/e$. In our simulations  the  sizes of the large and small openings,
$H$ and $h$, are kept fixed.}
}
\label{modelABC}
\end{figure}

We have conducted numerical experiments on channels of three different shapes, 
as illustrated in \ref{modelABC}.
Such choice is intended to capture the behavior of a large class of solid-state devices and
biological ionic channels, i.e. macromolecular pores that control the passage of ions across 
cell membranes. As shown in \ref{modelABC}, in system A the fluid mixture 
is confined to move between two  uniformly charged parallel plates. Such a geometry
is useful  to study the electroosmotic flow
and ionic current when the plate separation $H$ becomes of the order of the Debye length \cite{Kirby}. 
In system B  the surface charge of areal density $\Sigma$ covers only  two rectangular
regions of area ${\cal A}$ and longitudinal size $a$ on opposite  walls.
Finally, system C is obtained by considering a parallel slit  containing a narrowing of length $L$.
Such a  striction forms an open wedge, where the
planes are separated  by a distance $h$ at the small opening and $H$ at the large opening.
The wedge geometry recalls the shape of synthetic conical nanopores, which have been manufactured 
and studied in recent years  \cite{Siwy1,Siwy2, Cervera}  in order to understand the origin
of rectifying devices, as related to biological channels.

It is important to remark that, for models B and C, away from the charged region we do not explicitly 
model a completely macroscopic reservoir. Rather, the far away region acts as a buffer that allows
to recover electroneutrality sufficiently away from the charged region, as detailed in the following.
Given the studied geometries, it is of little usage to compute characteristic curves and compare them
directly with the experimental data obtained in electrolytic cells, 
as studied by Siwy and coworkers \cite{Siwy1,Siwy2}, whereas our systems are more alike series of
nano-devices. 

Throughout this paper we shall use lattice units defined in the following.
Time is measured in units of the discrete step $\Delta t=1$,
lengths are measured in units of lattice spacing, the charge $e$ and the mass $m$  are assumed to be unitary.
The thermal energy $k_BT$ is specified by fixing the thermal velocity whose value is
$v_T=\sqrt{k_B T/m}=1/\sqrt{3}$ in lattice units.
The dielectric constant is obtained by fixing the Bjerrum length, $l_B=e^2/(4\pi\epsilon k_B T)$, 
which is the distance at which the thermal energy $k_B T$ is equal
to the electrostatic energy between two unit charges.
We remind that in water under ambient
conditions and for two monovalent  ions this length is $ {\it l}_B \approx 0.7$ nm,
therefore, we set $l_B/\sigma = 4$.

We consider mixtures of hard spheres of equi-sized diameter $\sigma^{\alpha\alpha}\equiv \sigma=4$ in lattice units. The
packing fraction is fixed to ${\pi\over 6} \sum_\alpha \na (\sigma^{\alpha\alpha})^3 = 0.20$.
The surface charge density $\Sigma$ is always taken to be negative and with  typical values 
$\Sigma \sigma^2 / e = 10^{-3}$.
At initial time, the densities of the ions in the channel are set by fixing the
Debye length in the bulk:
\be
\lambda_D=\sqrt{\frac{\epsilon  k_B T}{2 e^2  n_{s}}}= \sqrt{\frac{1}{8\pi {\it l}_B   n_{s}}}  
\ee
where $n_s$ is the average density of a single ionic species in the bulk.
This relation is used to define the initial density of the minority ions (the coions) to $n^-(\rr)=n_s$.
Next, we impose a global electroneutrality condition in order to set the local density of counterions, as
\be
e\int_V(n^+(\rr) - n^-(\rr))  d\rr + \int_S \Sigma(\rr) d\rr = 0
\label{englobal}
\ee
where the first integral extends over the volume of the system and the second over its surface.
During the simulation, the number of particles of each species is fixed and, as time marches on,
the densities rearrange according to their intrinsic evolutions.
We fix the value at the Debye length to $\lambda_D/\sigma=2.75$, unless differently stated.

To fix the physical units, by stipulating that $\sigma=3\times 10^{-10} m$, 
we consider a mixture with number densities $n^0\sigma^3 =0.392$ and $n^\pm\sigma^3 = 0.001408$, 
corresponding to a $0.1\,M$ solution. 
The friction coefficient $\gamma$ for such mixture is about $0.5$ 
and the electric conductivity is $\sigma_{el}=8 \times \, 10^{-5}$.
Finally, the geometrical parameters of the pore are set to $H/\sigma=12.5$, $h/\sigma=2.75$ and $a/\sigma=10$.

In the following we compare results at finite packing fraction with results obtained 
in the absence of HS collisions, that is, at negligible packing fraction.
The former is useful to single out the role of molecular interactions
while the latter corresponds to the simple Bhatnagar-Gross-Krook approximation,
that is, the completely macroscopic picture.

Although the numerical method produces complete time dependent solutions,
we focus on studying stationary conditions for the sake of simplicity. 
For each simulated system, the convergence towards stationarity is fast ($\sim2 000$ iterations)
and in general very stable. 
In the near future, we plan to explore time dependent phenomena, as in alternate electric field conditions. 

\begin{figure}[htb]
\includegraphics[bb=30bp 30bp 440bp 420bp,clip,width=10.5cm, keepaspectratio]{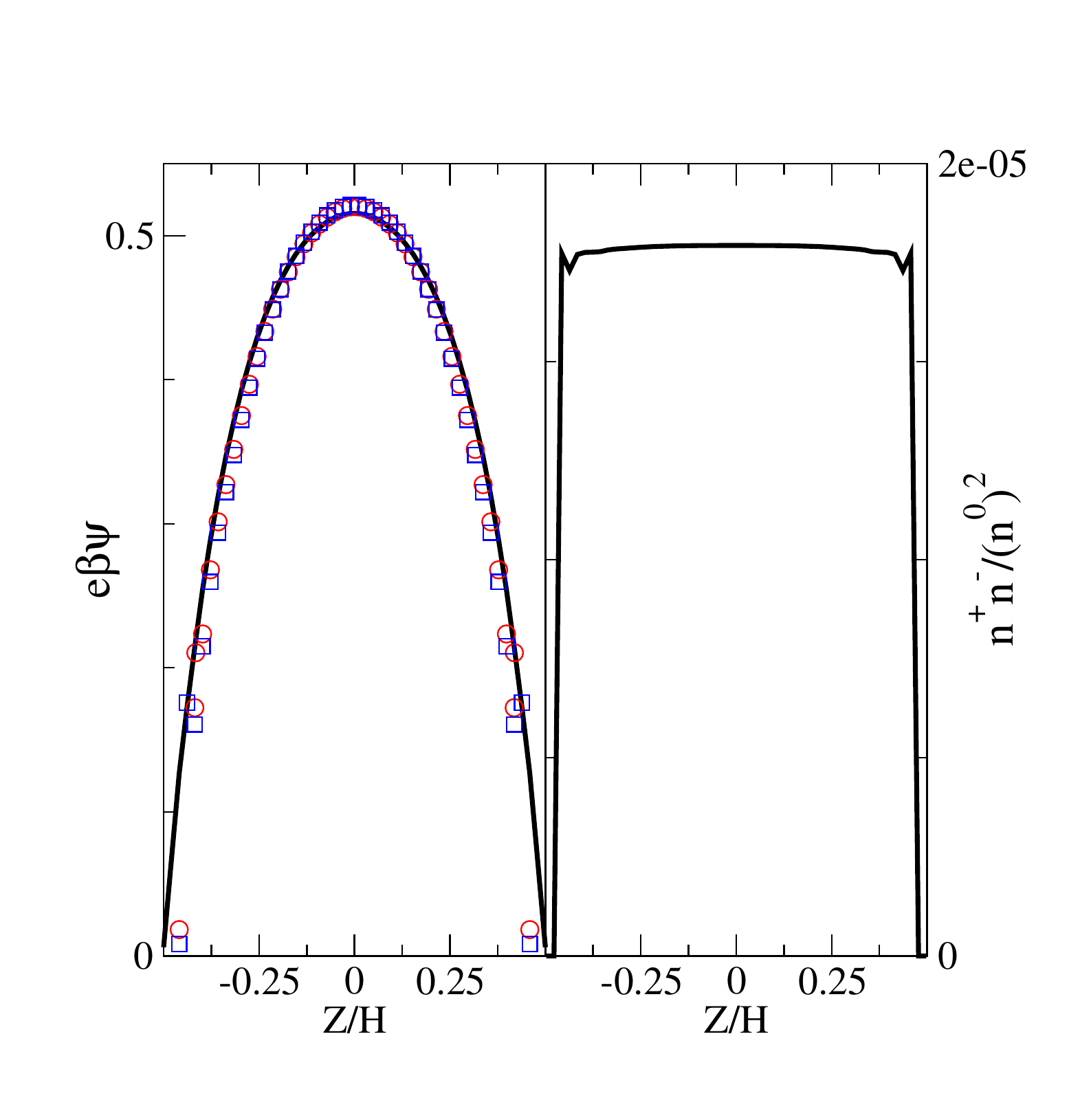} \\
\includegraphics[bb=30bp 30bp 440bp 420bp,clip,width=10.5cm, keepaspectratio]{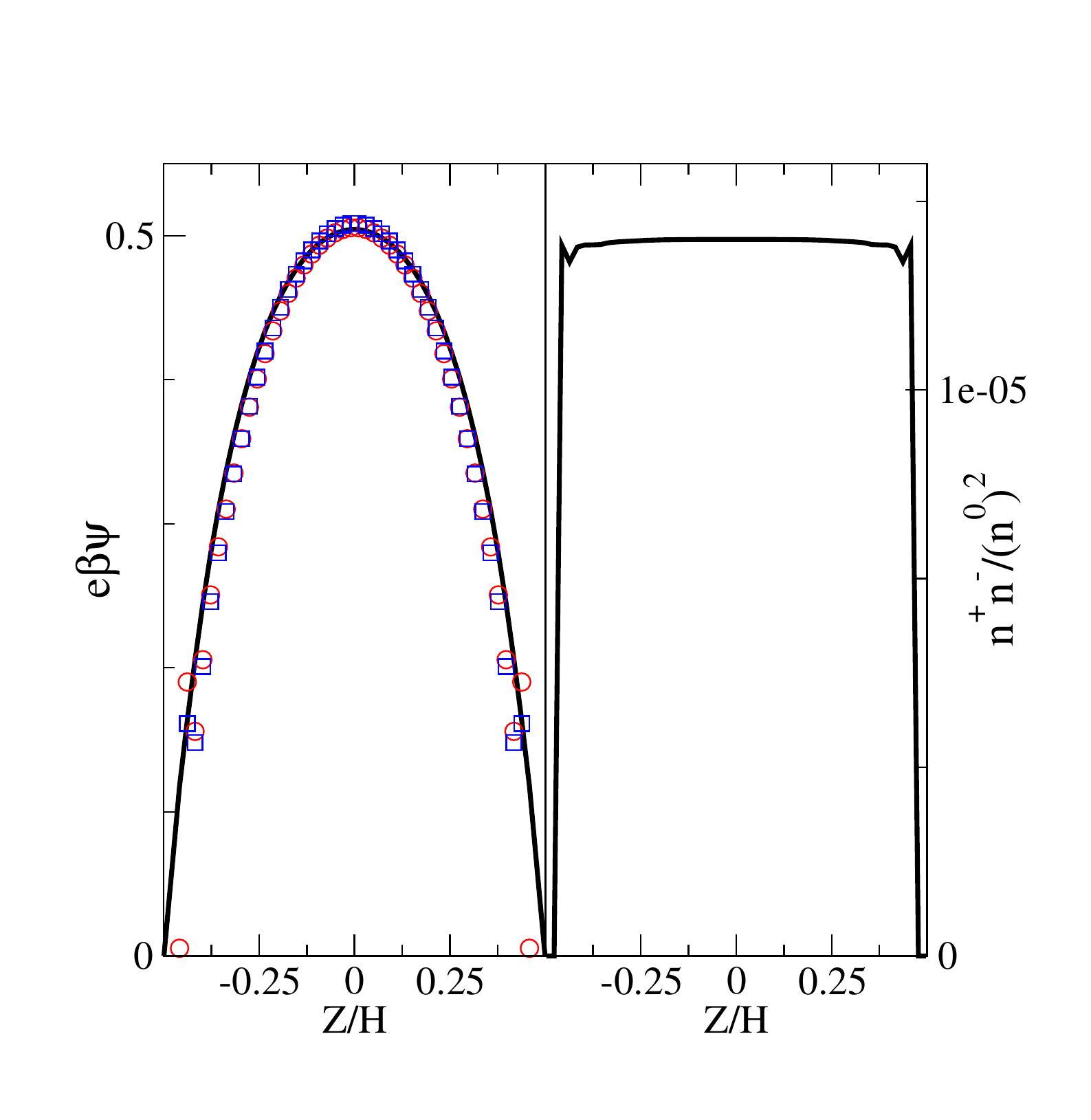} \\
\caption{Validation of the equilibrium conditions \ref{validaeq1} and 
\ref{validaeq2} for models A (upper) and B (lower panel).
The left panels report the transversal profiles of the scaled potential $ e \psi/k_B T$ (solid curves) and
$A + \ln (n^0/n^+)$ (circles) and $B - \ln (n^0/n^-)$ (squares), with $A$ and $B$ constants. 
The right panels report the transversal profile of the product $n^+ n^-/(n^0)^2$. 
All profiles are taken at $x=L/2$.
}
\label{densityratiochecksmodAB}
\end{figure}

\begin{figure}[htb]
\includegraphics[bb=30bp 30bp 440bp 420bp,clip,width=12.5cm, keepaspectratio]{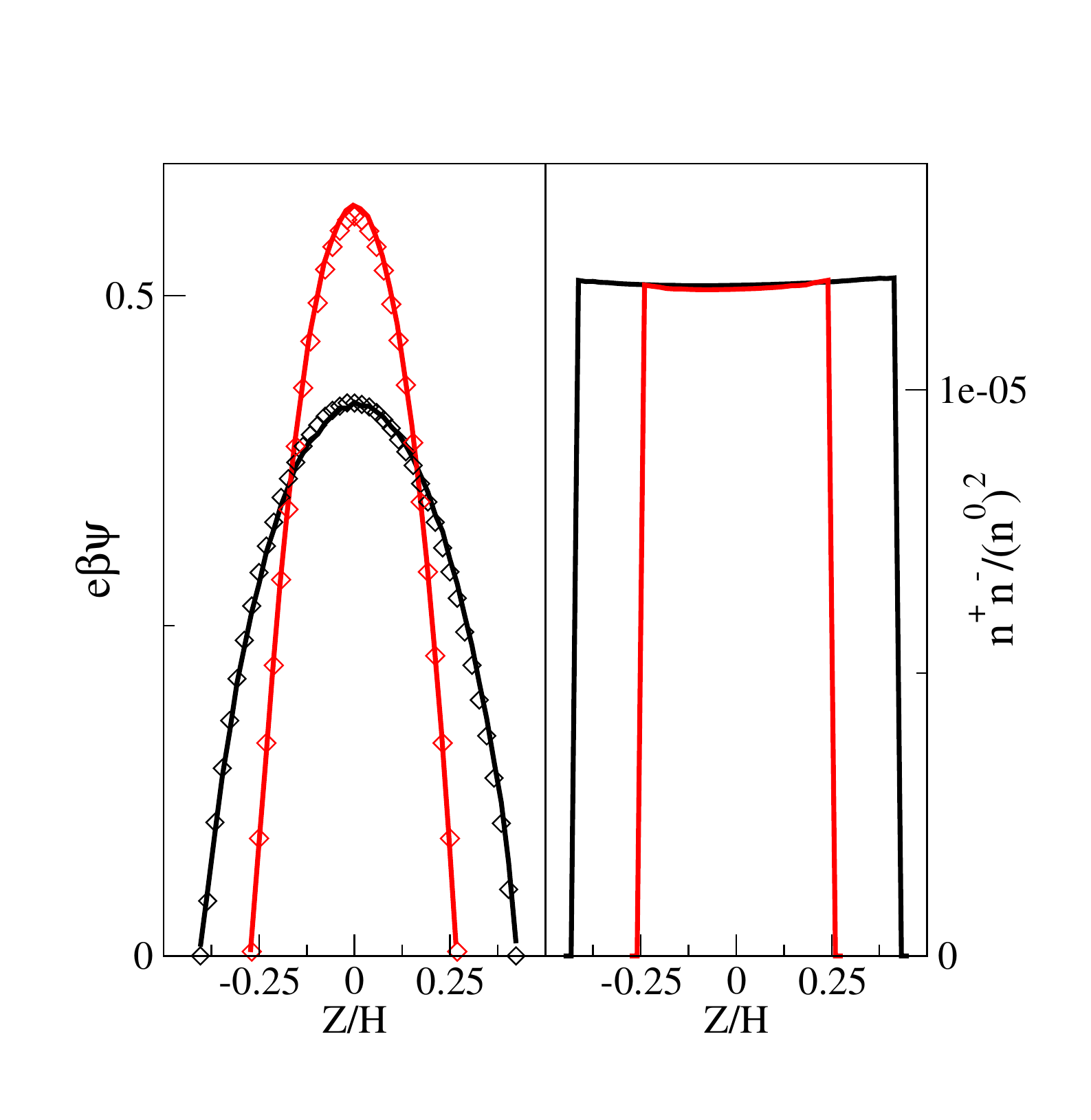}
\caption{Same as \ref{densityratiochecksmodAB} for model C.
The left panels report the transversal profiles of the scaled potential $ e \psi/k_B T$ (solid curves) and
$A + \ln (n^0/n^+)$ \ttr{(diamond symbols)}, with $A$ a constant. 
The right panels report the transversal profile of the product $n^+ n^-/(n^0)^2$. 
The curves are transversal profiles taken at the pore larger mouth \ttr{(black line and symbols)}
and at midpoint of the 
channel \ttr{(red line and symbols)}.  
}
\label{densityratiochecksmodC}
\end{figure}

As a preliminary test, we validate the equilibrium properties of the system since
in the absence of ionic currents, the theory predicts a simple local relation between the
species densities and the electric potential obtained as immediate consequences of \ref{jolyrelation}
and reading
\be
{n^+ (\rr,t) \over n^-(\rr,t)} = {C^+ \over C^-} \exp(-\frac{2 e z^\pm \psi(\rr)}{k_B T})
\label{validaeq1}
\ee
and
\be
{n^+ (\rr,t) n^-(\rr,t) \over n^2_0(\rr,t)} = C^+ C^- 
\label{validaeq2}
\ee
that is, the ratio of the density profile probes the electrostatic field $\psi(\rr,t)$, 
whereas the product of the two profiles should be everywhere proportional to the square of the density of the solvent.
These ratios are  displayed in \ref{densityratiochecksmodAB} and \ref{densityratiochecksmodC} and the agreement is quite good for both tests. 
Joly and co-workers have performed a similar test in their study \cite{Joly}, but in contrast to our method,
they made the comparison by utilizing the density profiles
obtained from MD simulation of a ternary charged mixture. In the present calculation instead both the
electric potential and the density profiles are obtained from the same numerical calculation in a self-consistent
fashion.

\begin{figure}[htb]
\includegraphics[clip=true,width=12.0cm, keepaspectratio]{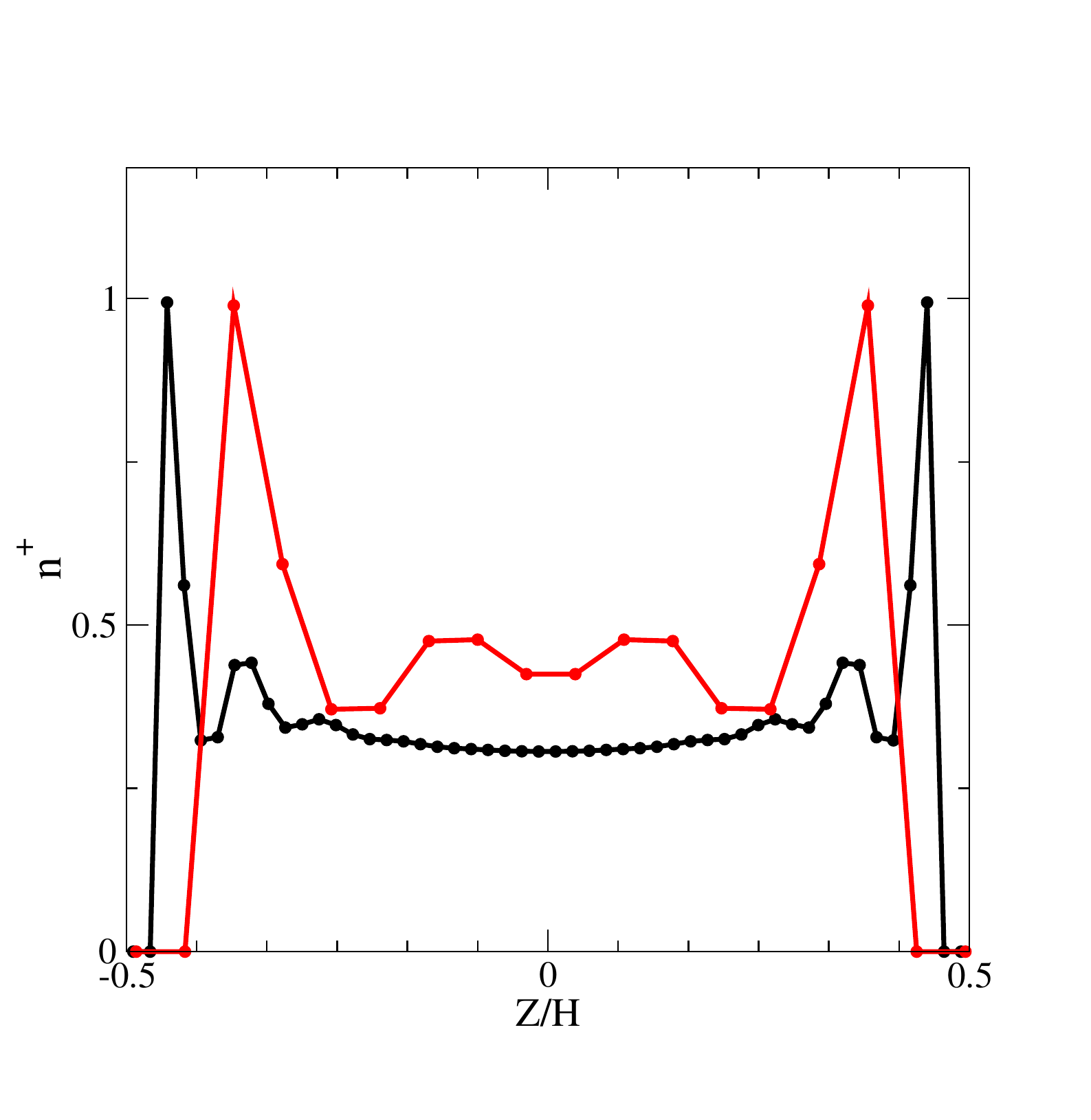}
\caption{Model A: density profile \ttr{of the counterions} obtained with the full electrostatic interactions 
(\ref{Poisson}) and subtracted by the contribution of \ref{poissoncorrection}, for packing fraction $0.20$, 
$\Sigma=-0.0003$ and for channel width $H/\sigma=12.5$ (black line and red symbols) and $H/\sigma=4.5$
(red line and red symbols). The profiles have been rescaled by their maximum value for the sake of readability
and are virtually indistinguishable.}
\label{comparecoulomb}
\end{figure}

\tr{The usage of the full Poisson equation in connection with a hard sphere fluid
disregards the effect of charge correlations \cite{corry}. This effect has been included 
by some authors \cite{Nilson} using the Mean spherical approximation (MSA) 
represention of the direct pair correlation function.
At the MSA level, the exclusion region of hard spheres is removed from the source term 
of \ref{Poisson}.
We have thus considered this correction to the electrostatic forces by removing the contribution
\begin{equation}
- {1\over \epsilon} \int_{|\rr-\rr'|<\sigma} d\rr' \rho_e(\rr') \frac{(\rr'-\rr)}{|\rr-\rr'|^3} = 
- {1\over \epsilon} \int_{|\rr-\rr'|<\sigma} d\rr' (\rho_e(\rr') - \rho_e(\rr)) \frac{(\rr'-\rr)}{|\rr-\rr'|^3}
\label{poissoncorrection}
\end{equation}
from the interaction. Notice that the second part of \ref{poissoncorrection} 
is a simple rewriting of the left hand side that
regularizes the integral and avoids numerical overflows.
The profiles of the counterions 
with and without the correction \ref{poissoncorrection} are reported in \ref{comparecoulomb}. 
It is apparent that at the molarities of interest as used in the present work, 
the effect of interactions arising from inner cores is very small for large and narrow pores
and can be safely ignored in the simulations.
}

\begin{figure}[htb]
\includegraphics[bb=0bp 30bp 440bp 420bp,clip,width=10.5cm, keepaspectratio]{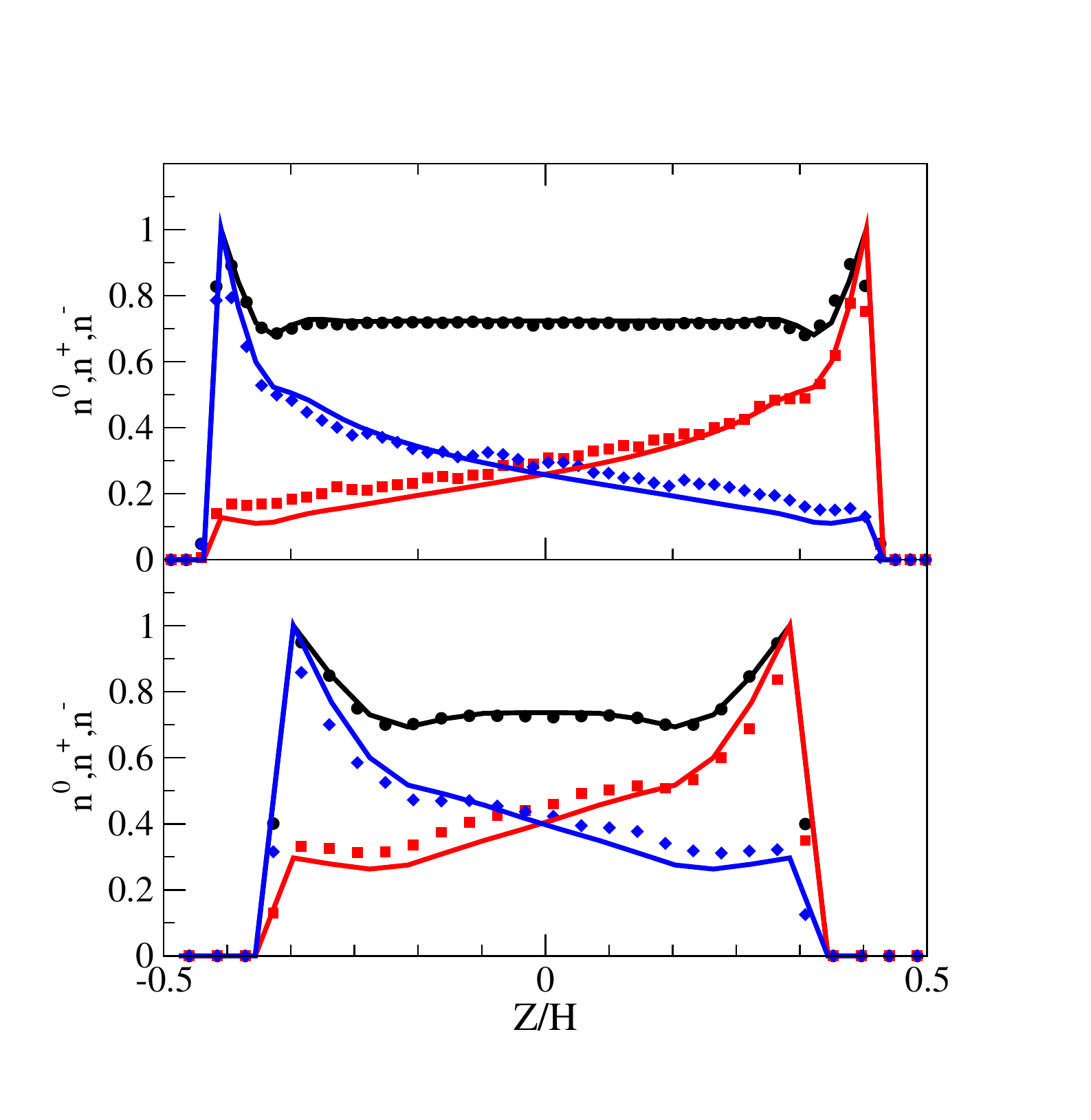} \\
\includegraphics[bb=0bp 30bp 440bp 420bp,clip,width=10.5cm, keepaspectratio]{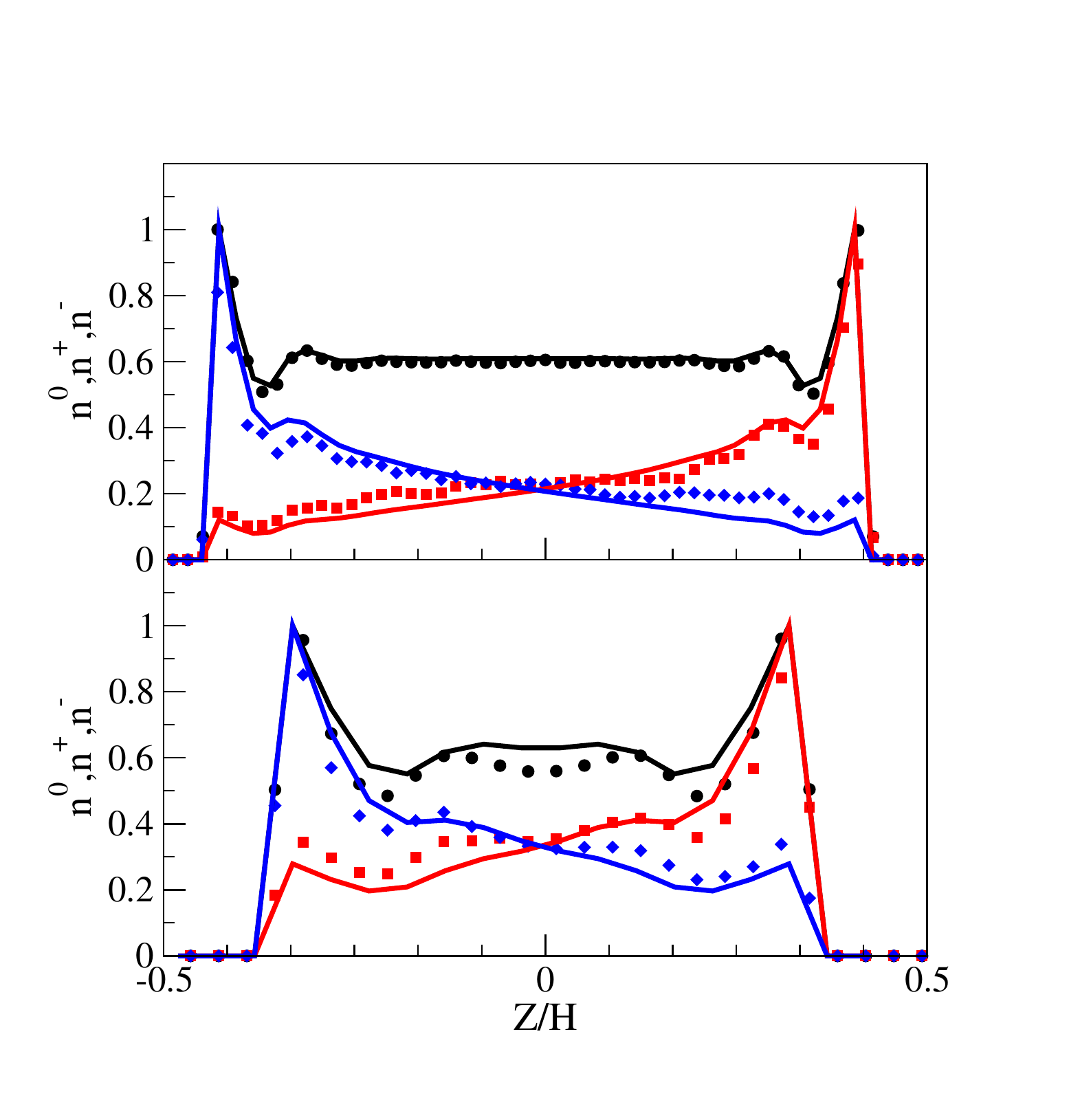}
\caption{Comparison of density profiles as obtained from the electrokinetic method and Molecular Dynamic simulations
for an equivalent soft sphere system. Simulation details are reported in the text.
All profiles have been rescaled by its maximum value for the sake of readability.
}
\label{comparemd}
\end{figure}

\tr{
A further test of the quality of the theoretical framework comes from comparing the equilibrium
density profiles with the results of Molecular Dynamics 
simulations of a ternary mixture confined between two parallel slabs of 
}\tr{
opposite surface charges, that is, a slight variation of Model A. 
In the MD simulations the excluded volume interactions are modelled via a Lennard-Jones potential
\begin{equation}
V_{LJ}(r_{ij}) = 4\epsilon 
\left[ \left( \frac{\sigma}{r_{ij}} \right)^{12} -  \left( \frac{\sigma}{r_{ij}} \right)^{6} \right]
\end{equation}
truncated at the cutoff, so that $V_{LJ}=0$ for $r_{ij}>2^{1/6} \sigma$.
The particles are repelled from the slab walls by the same type of interaction.
In addition, the electrostatic forces are computed via the Ewald summation method for a periodic system
\cite{frenkelsmit}.
The MD simulations are performed at constant volume and temperature, with parameters of
$\epsilon / k_BT = 2.5$ and $\sigma = 2$.
The comparison between the electrokinetic and MD results is done by determining the exclusion radius of the 
soft spheres used in MD, that is, by finding the effective distance $r^*$ such that 
$V_{LJ}(r^*) = k_B T$. An analogous procedure is applied to determine the effective volume of the confining slab
and, in this way, to match the densities of each component between the electrokinetic and MD simulations. 
The MD profiles are accumulated over $10^{7}$ timesteps and for two different slabs of width
$H/\sigma=10$ and $4.5$, filled with $\simeq 2000$ and $\simeq 4000$ particles resulting in
packing fractions of $0.10$ and $0.20$, respectively. 
In order to accumulate sufficient statistics we perform the MD simulations at sufficiently high molarities
corresponding to  densities of $n^\pm=2.0\times 10^{-5}$.
The electrokinetic and MD profiles reported in \ref{comparemd} are in excellent agreement, 
in particular for the larger slab width. The quality of results does not
seem to deteriorate with the packing fraction considered here.
The above results are particularly reassuring in view of the small molarity accessible by
the electrokinetic methodology, and overall provides good confidence on the simulation framework.
}

\begin{figure}[htb]
\includegraphics[clip=true,width=16.0cm, keepaspectratio]{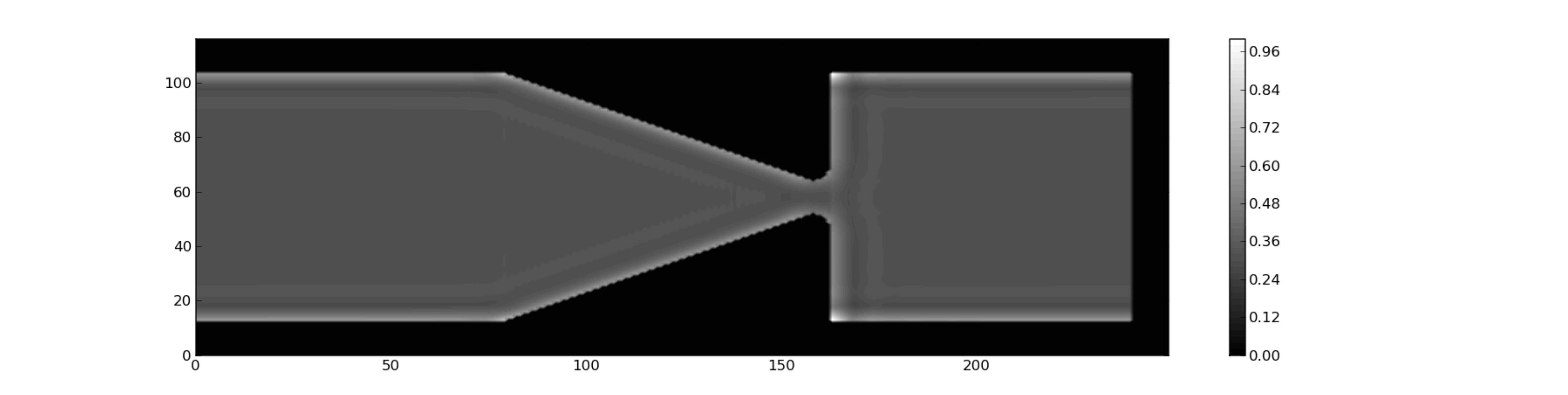}
\includegraphics[clip=true,width=16.0cm, keepaspectratio]{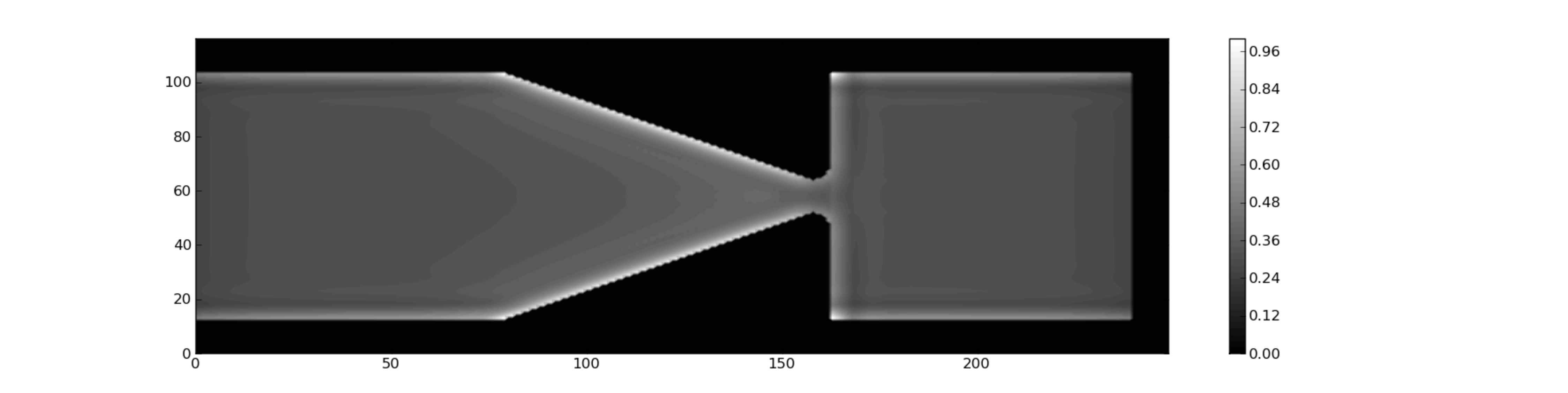}
\includegraphics[clip=true,width=16.0cm, keepaspectratio]{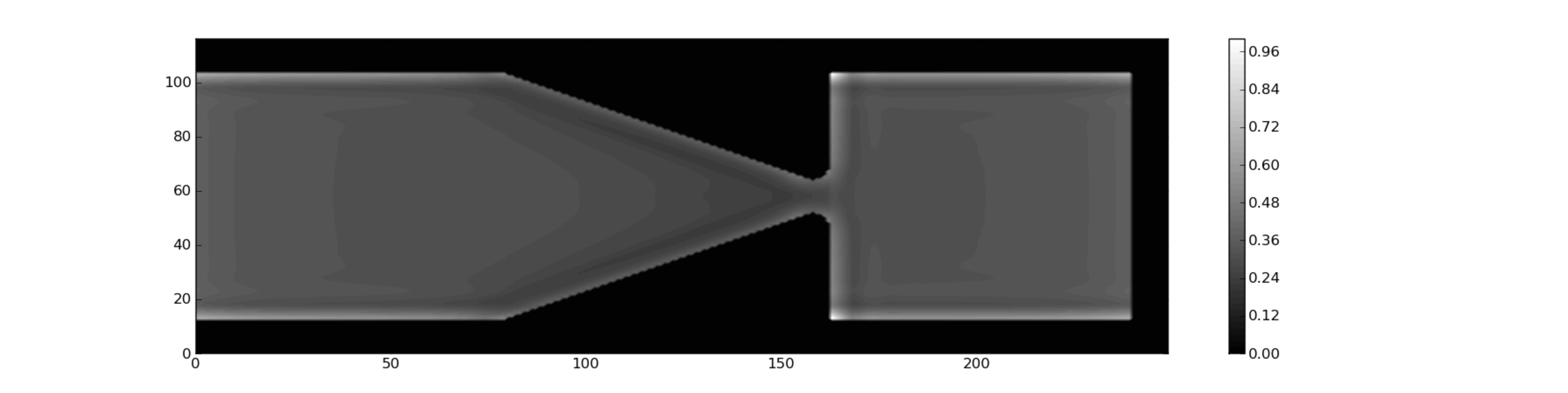}
\includegraphics[clip=true,width=16.0cm, keepaspectratio]{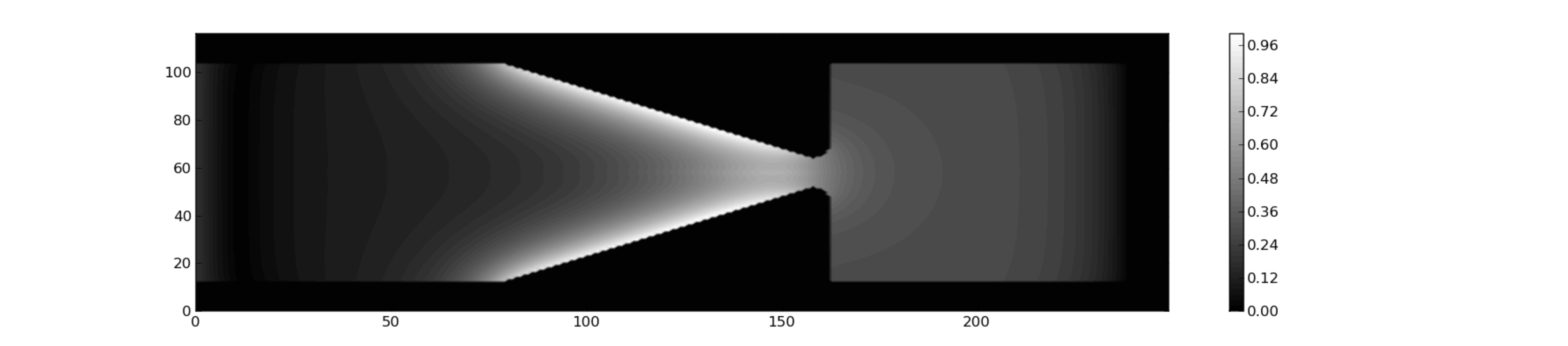}
\includegraphics[clip=true,width=16.0cm, keepaspectratio]{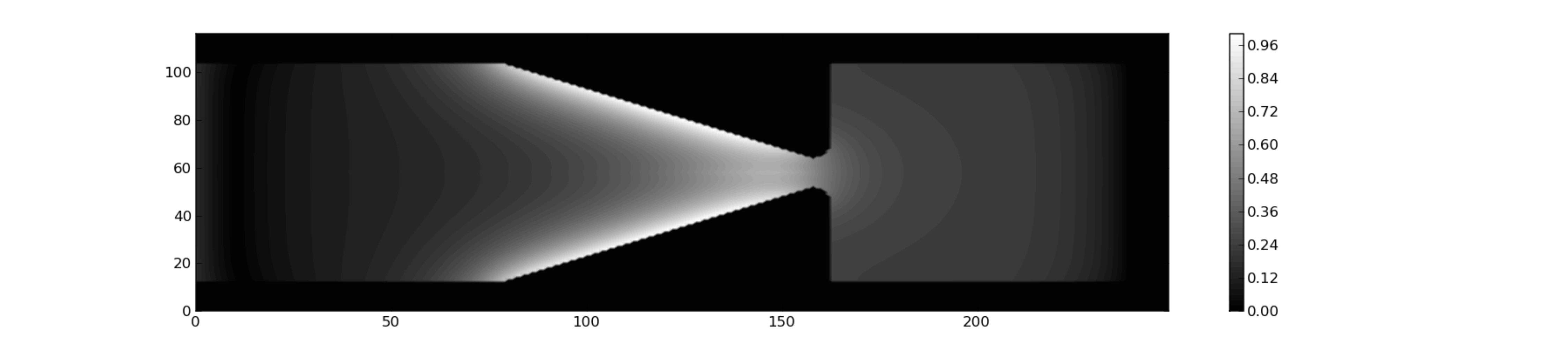}
\caption{Model C: contours of scalar fields for the case with $\Sigma=-0.002$ and voltage $0.3\,k_BT$. 
Top to bottom: density of the neutral species, density of counterions, density of coions and
electrostatic potential. The lowest panel is the electrostatic potential for the equivalent system
without hard sphere forces.}
\label{contour}
\end{figure}

{One the of reasons for studying wedged channels having a nanometric orifice is the
current rectification properties of this type of devices. In the past, such behavior has been ascribed to a ratchet
mechanism arising from the joint effect of the asymmetric electrostatic potential within the pore region and
departures from thermal equilibrium \cite{Siwy1,Siwy2}. Alternatively, it has been proposed that rectification arises from the
inhomogeneous conductivity near the orifice that would produce asymmetry of fluxes of counterions and coions \cite{WOERMANN}.
In this context, a third cause of rectification
could be due to local charge accumulation in the wedge pore region, that is, an imperfect Donnan compensation
of surface charges. It is thus interesting to look} 
at the highly non-trivial distribution of neutral and charged species, as displayed in \ref{contour} 
for the contour plots for the density of species
in model C for a surface charge of $\Sigma=-0.002$.
The plots show that the several walls present in the wedge geometry, strongly modulate the density profiles
of each species as the effect of layering, and this layering overlaps with the double layer organization.
We further notice that the concave corners act as strong accumulation points for the species, similarly to what
shown some time ago by Dietrich and co-workers \cite{Dietrich}. 
The reason for the accumulations is completely entropic and is such that the concave corners effectively 
attract particles, while convex corners exert repulsive forces.
\ref{contour} reports the electric potential in the same geometry at zero and finite packing fraction
and, surprisingly, the self-consistent solution for the potential does not depend sensibly on the presence 
of hard sphere interactions.

\begin{figure}[htb]
\includegraphics[bb= 0bp 30bp 440bp 420bp,clip,width=10.0cm, keepaspectratio]{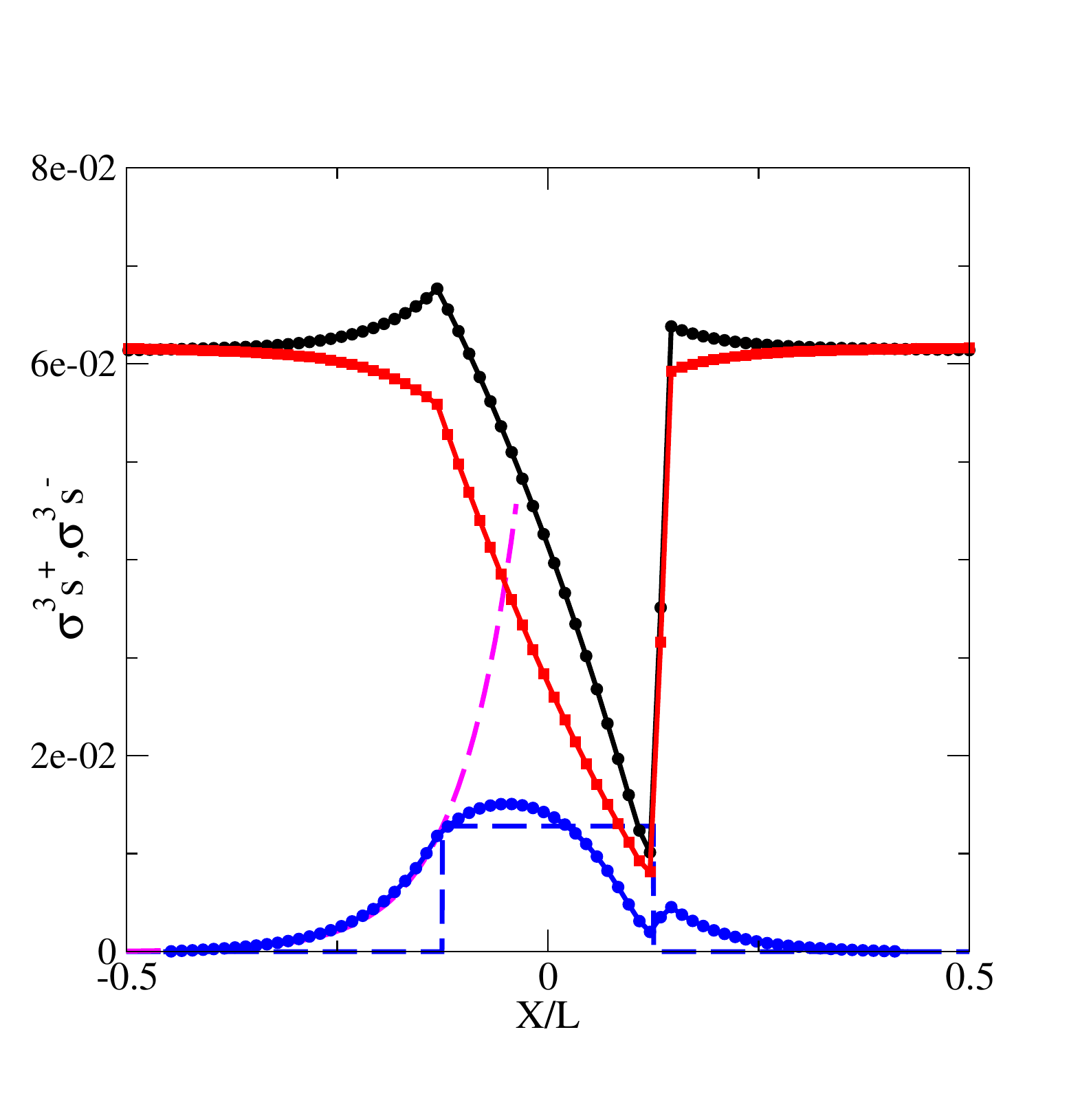}
\includegraphics[bb= 0bp 30bp 440bp 420bp,clip,width=10.0cm, keepaspectratio]{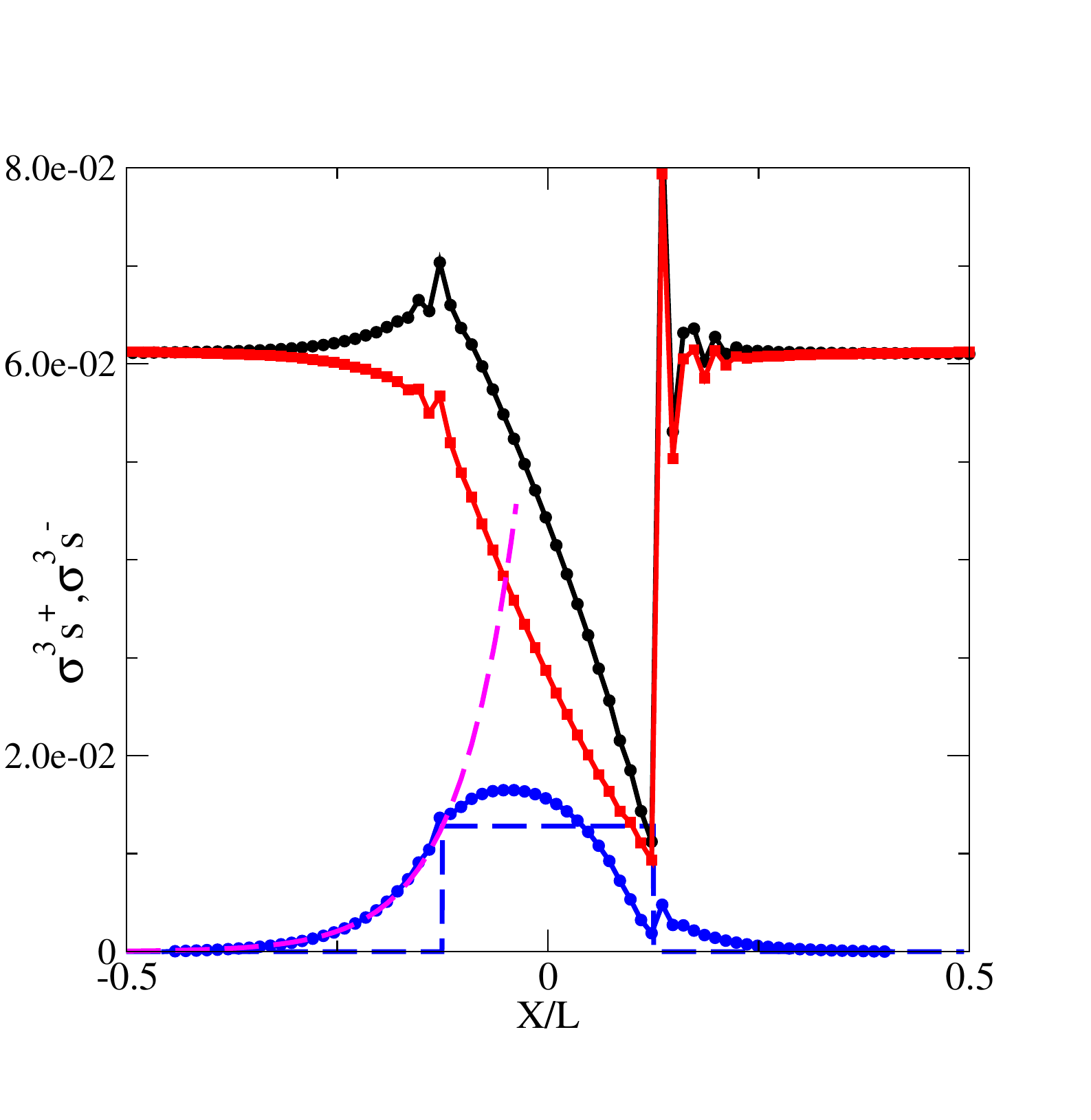}
\caption{
Model C: transversally averaged density profiles of the two ionic species with (upper panel) and without (lower panel)
HS collisions.
The black curve represents the counterion profile $s^+(x)$,
the red curve the coion profile $s^-$, and the blue curve their difference (the local charge).
The latter has to be compared with the surface charge density profile (blue stepwise dashed curve).
\tg{Finally, as a guide to to the eye} the exponential decay with a characteristic
Debye length, $\lambda_D$, is reported as a violet dashed curve.
}
\label{fettonicarica}
\end{figure}

We next evaluate the charge count along the $x$ direction and in the absence  of external voltage for model C.
The charge count is defined as $s^\pm(x) \equiv {1 \over S} \int n^\pm(x,y,z) dy dz$,  where $S$ being the sectional area,
which can be assimilated to a local surface charge of each species. 
The local excess charge $\Delta s\equiv (s^+ - s^-)$ should partially compensate for the surface charge 
in the wedge region. 
\ref{fettonicarica} shows several interesting features that strongly depend on packing fraction.
At first, we notice the appearance of oscillations in the profiles induced by the packing effects 
which are more pronounced near the narrow exit of the wedge.
More importantly, the plots show that the compensation is imperfect, within the wedge region $\Delta s$
stays around $-\Sigma$ and displays strong similarities between the zero and finite packing fraction cases. 

\begin{figure}[htb]
\includegraphics[bb=0bp 30bp 440bp 420bp,clip,width=10.5cm, keepaspectratio]{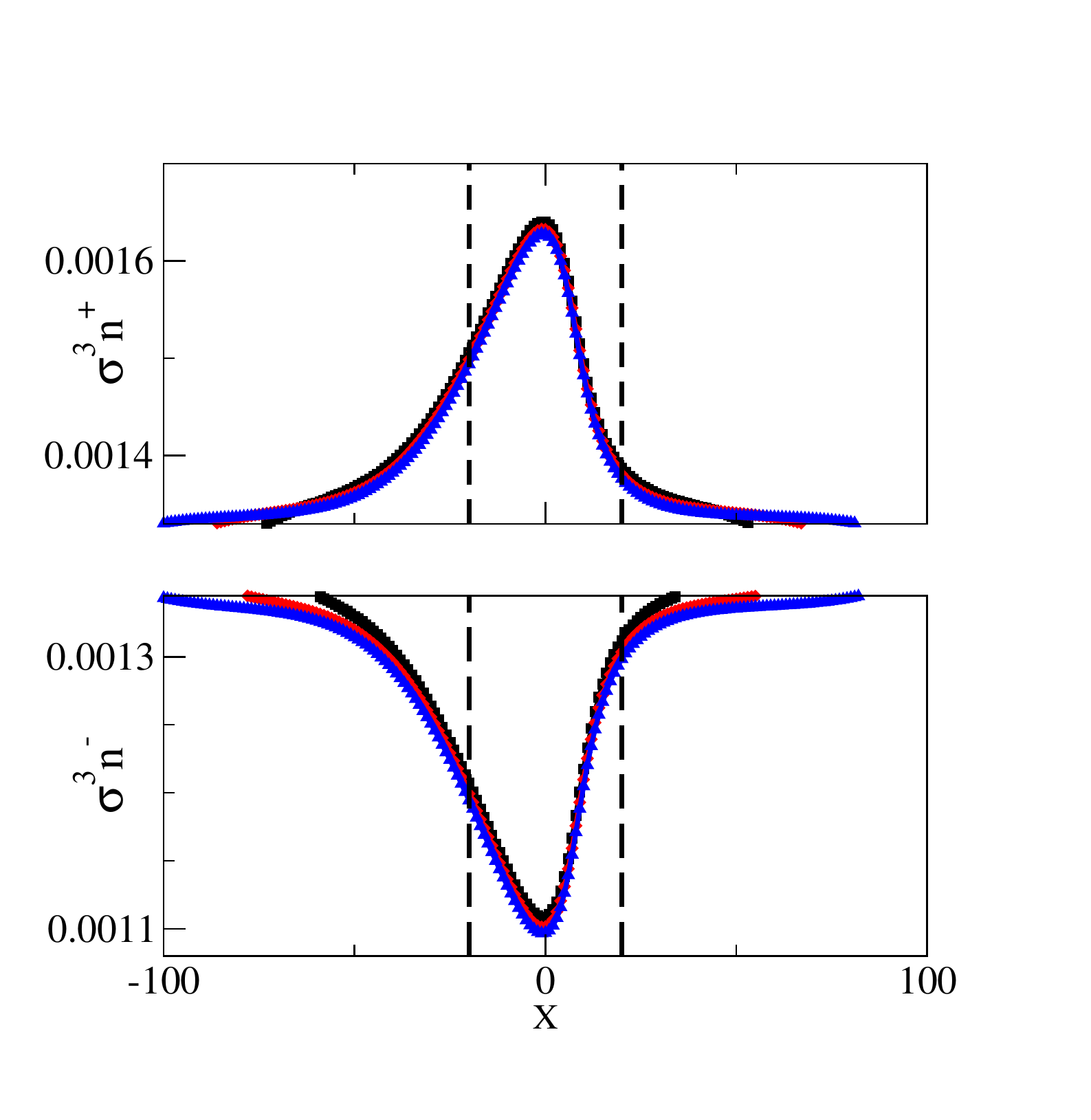}
\includegraphics[bb=0bp 30bp 440bp 420bp,clip,width=10.5cm, keepaspectratio]{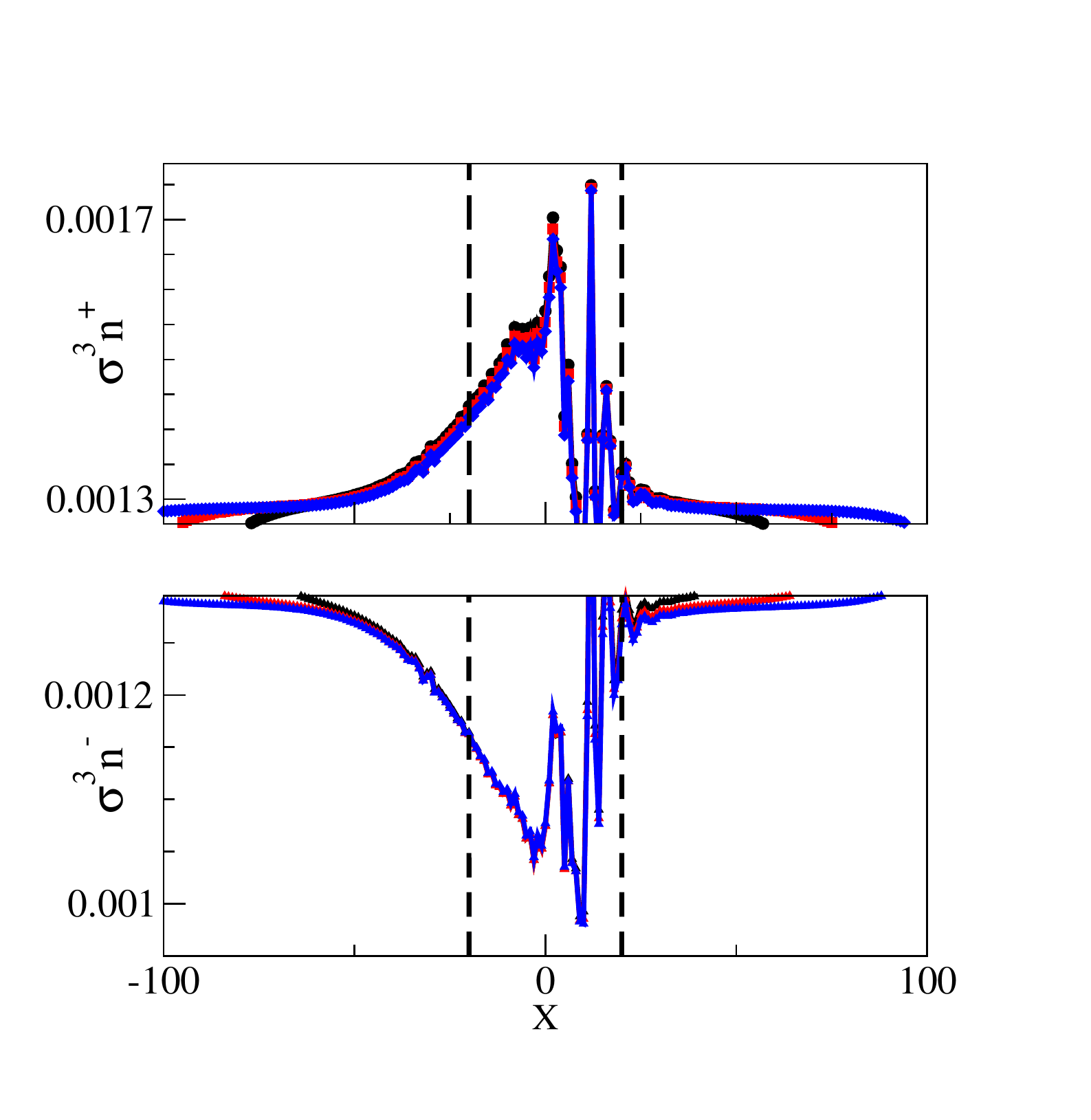}
\caption{Model C: density profiles along the channel centerline 
of counterions (upper panel) and coions (lower panel) without (left panel)
and with (right panel) hard sphere collisions. Profiles for different channel lengths are shown: 
$L=120$ (black), $L=160$ (red) and $L=200$ (blue) in lattice units. 
Data are for $\Sigma\sigma^2/e=-0.0032$ and zero voltage.  The dashed lines indicate the pore region.}
\label{figsizeeffect}
\end{figure}

\ref{fettonicarica} reveals how the molecular system departs from the assumption of local charge neutrality 
often utilized in electrochemical studies. Only away from the pore region the excess charge decays to zero,
that is, it restores electroneutrality. 
The longitudinal decay length is given by $\lambda_D$ on the side of the larger
wedge mouth and shows a slightly more complex behavior on the other side. 
It is an interesting fact that the Debye length controls the longitudinal restoration of
electroneutrality, perhaps not completely unexpected. 
In practical calculations, it is however important to determine the finite size effects, 
such as in the periodic system studied here. \ref{figsizeeffect} shows that
already channels of size $L/\lambda_D>8$ are sufficiently long to show a good convergence
of the charge distribution. For our simulations, we have chosen $L/\sigma=40$ as the reference
channel extensions to obtain satisfactorily converged results.

\begin{figure}[htb]
\includegraphics[bb=0bp 30bp 440bp 420bp,clip,width=15.0cm, keepaspectratio]{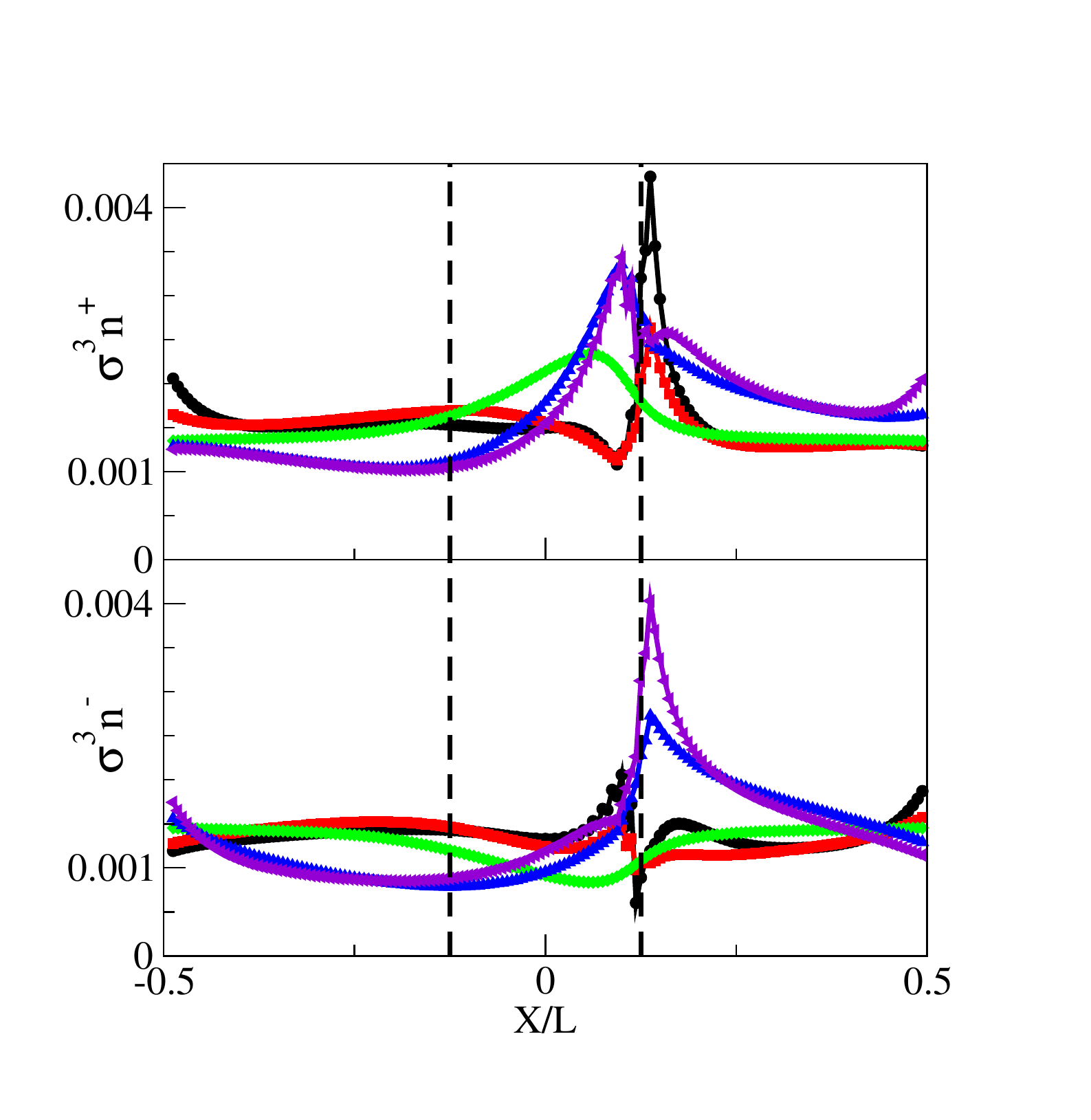}
\caption{Model C: density profiles along the channel centerline 
of counterions (upper panel) and coions (lower panel) 
in the presence of hard sphere collisions. Data for voltages $K/k_B T=-60$ (black), $-30$ (red), $0$ (green), 
$30$ (blue) and $60$ (violet), taken at $\Sigma \sigma^2 / e=-0.016$ and $L/\sigma=40$. 
The dashed lines indicate the limits of the pore region.}
\label{figcenterlinecharge}
\end{figure}

Once the voltage acts on the system, the charge unbalance modulates the charge organization
in highly non-trivial ways. 
The charge distribution along the channel centerline of the wedged geometry is shown in \ref{figcenterlinecharge}
for different voltages. Counterions tend to accumulate 
and coions deplete in the wedge region for positive voltages. The situation reverses by
reversing the bias and, under these conditions, we observe a strong charge piling up on the small mouth side.

\begin{figure}[htb]
\includegraphics[clip=true,width=14.0cm, keepaspectratio]{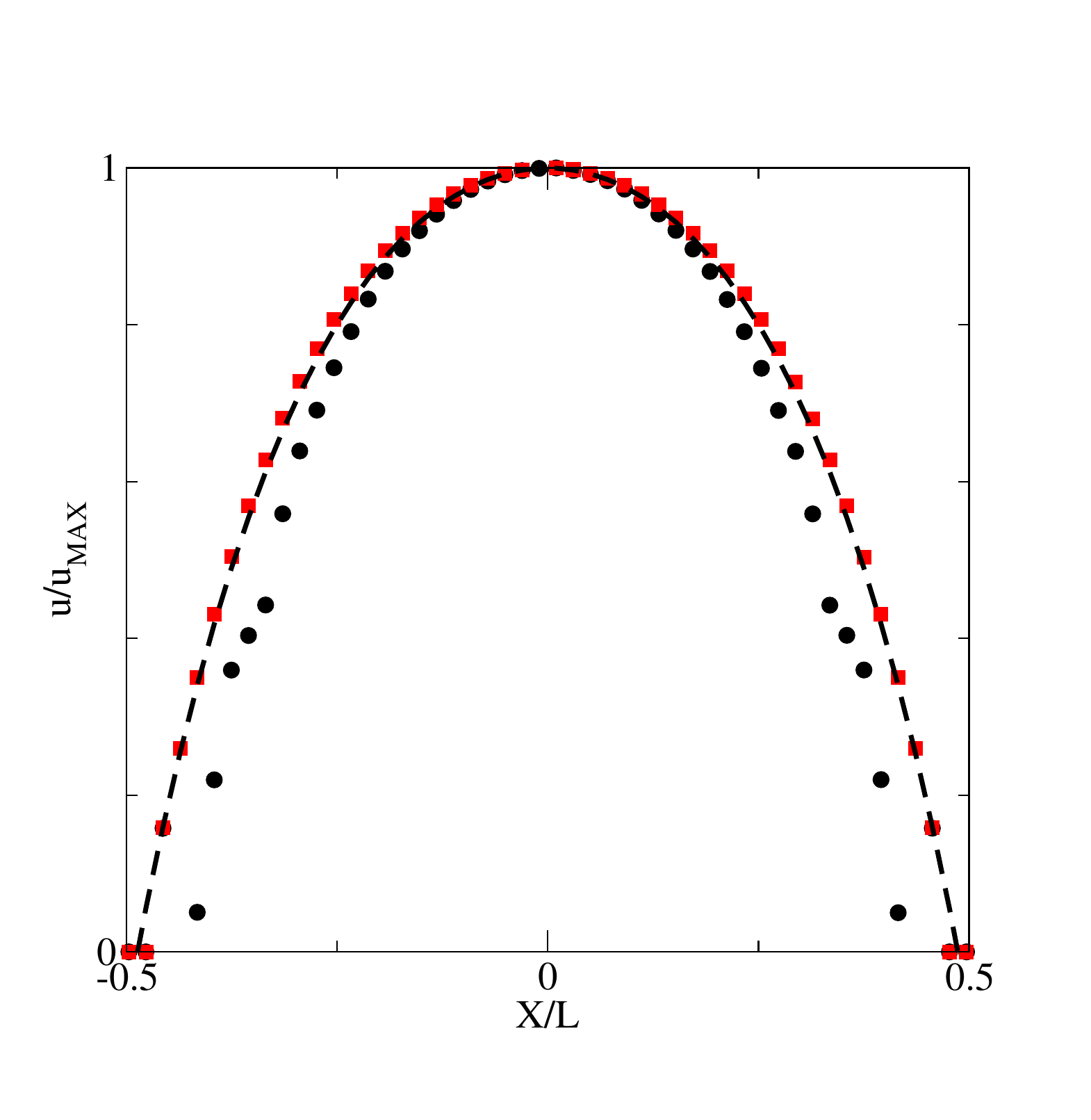}
\caption{Model A. Barycentric velocity divided by its maximum value.
The profile refers to an electrolyte mixture under a voltage $V/k_B T= 30$ in the absence (squares)
and in the presence of  HS collisions (circles).
The Smoluchowski analytical solution is also displayed (dashed curve).}
\label{Smoluchowski}
\end{figure}

In \ref{Smoluchowski} we turn our attention to the
mass transport process and display the barycentric velocity profile, $u_z(x)$,  of the fluid in the $x$ direction 
in the case of model A. For the sake of comparison we also show
the prediction of the  Smoluchowski analytical  formula \ref{vprofile}, obtained by
calculating the electrostatic potential $\psi(z)$ within the Debye-H\"uckel linear approximation.
We observe that the agreement between this formula and the results relative to the BGK system is very good, whereas 
the system with excluded volume interactions displays smaller velocities near the confining walls due to the larger value of the viscosity.

\begin{figure}[htb]
\includegraphics[bb=0bp 30bp 440bp 420bp,clip,width=10.5cm, keepaspectratio]{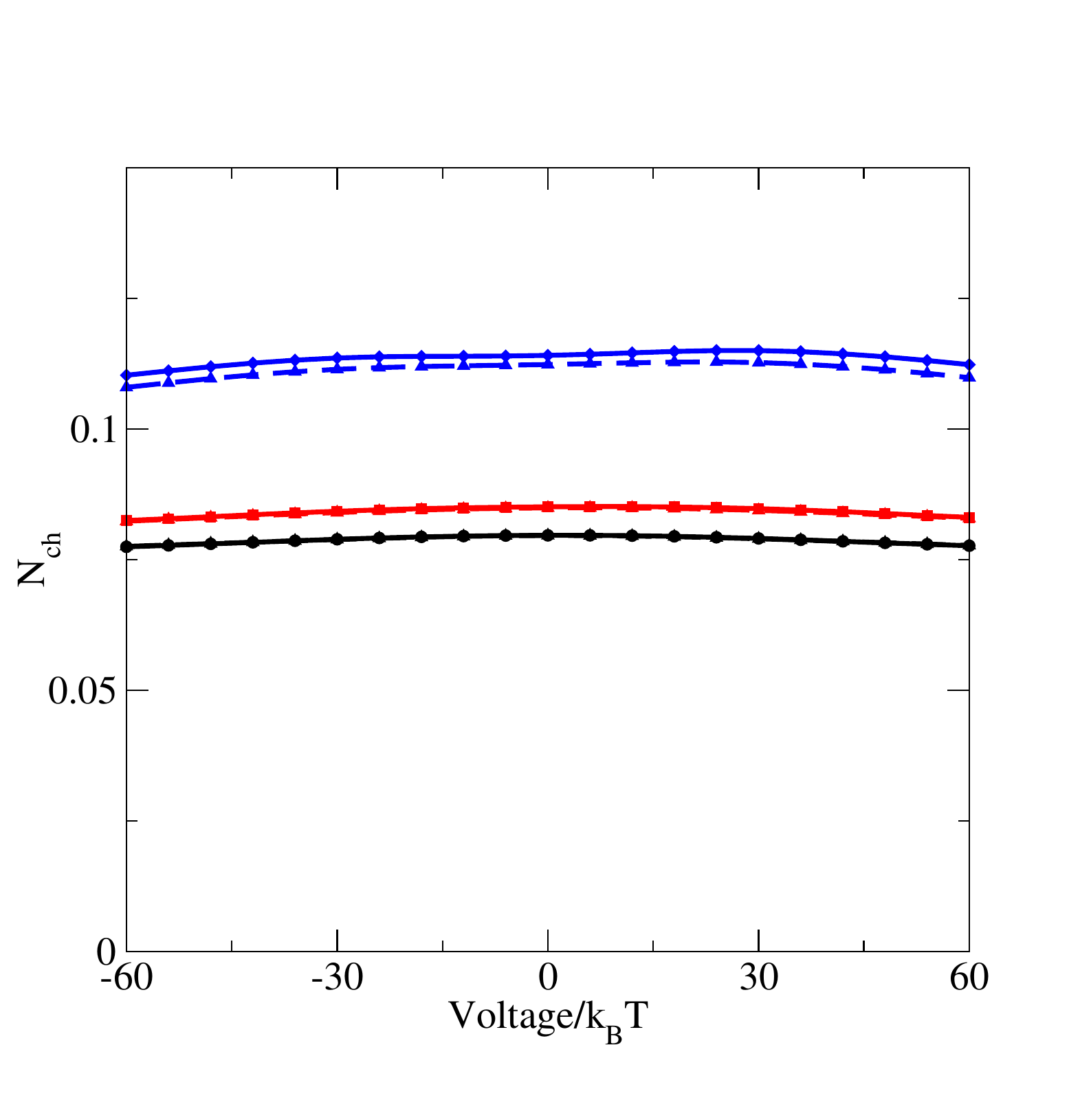} \\
\includegraphics[bb=0bp 30bp 440bp 420bp,clip,width=10.5cm, keepaspectratio]{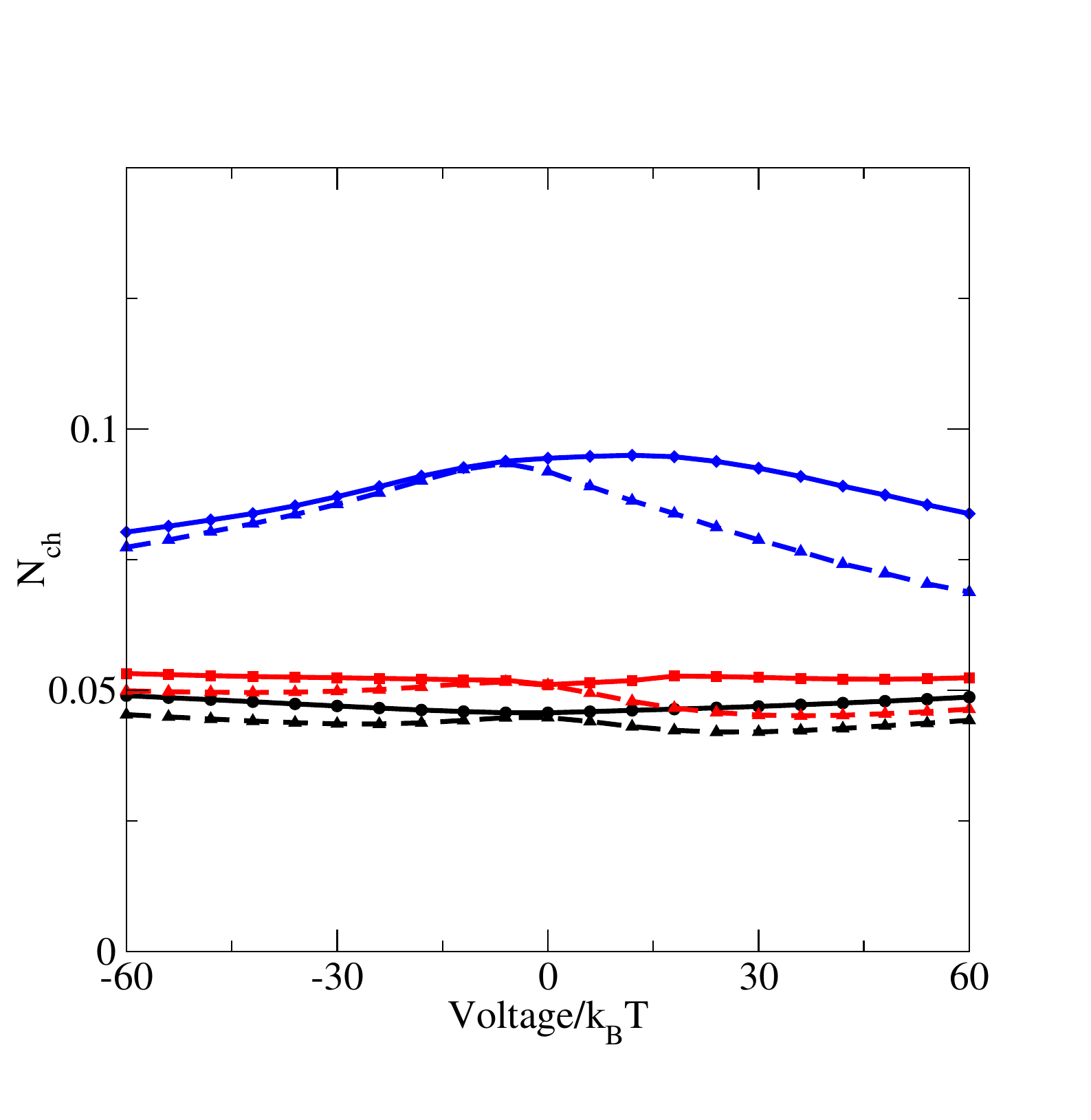} \\
\caption{Model C. Number of charge carriers in the pore region $N_{ch}=\int_{V_{pore}} d^3 r [n^+(r) + n^-(r)]$  
for model B (upper panel) and model C (lower panel). The curves correspond to
$\Sigma\sigma^2/e=-0.0016$ (circles), $-0.0048$ (squares), and $-0.016$ (triangles), respectively.
Solid lines correspond are in the presence of HS collisions while the dashed lines are without HS collisions
}
\label{figlocaloccupancy}
\end{figure}

{
Charge transport in the system is mostly due to ionic conduction rather than due to electro-osmotic transport.
We therefore compute the number of mobile charge carriers in the pore region.
As expected for model B, \ref{figlocaloccupancy} shows symmetric curves upon voltage inversion
and rather flat curves. On the other hand, for the wedge system, the curves lose their symmetry
and become more pronounced with the voltage. 
Only at the largest value of the surface charge
density, one observes an appreciable deviation between the two channel shapes.
In the lower panel of \ref{figlocaloccupancy}, the effect of the wedged geometry determines a non flat dependence 
of the number of the mobile charge carriers on the
applied voltage. Also we notice that this number is lower in the presence rather than in the absence of HS collisions, 
since the excluded volume interaction hinders the pile up of charges in the wedged region. 
In any case, the undercharging of the pore at non-zero voltage is rather limited ($<20\%$).
By the same token, the imperfect neutralization of the wedged region
decreases the Drude-Lorentz conductivity for both positive and negative voltages, as compared to the zero voltage condition.
The observed modest asymmetry in local charging also reflects in differences in conductivity for forward vs backward bias.
However, rectification usually shows up as a much stronger effect \cite{Siwy1,Siwy2}, and therefore we exclude the fact that
it is due to modulations of the occupation of the wedged region by mobile charge carriers.
}



\begin{figure}[htb]
\includegraphics[clip=true,width=12.0cm, keepaspectratio]{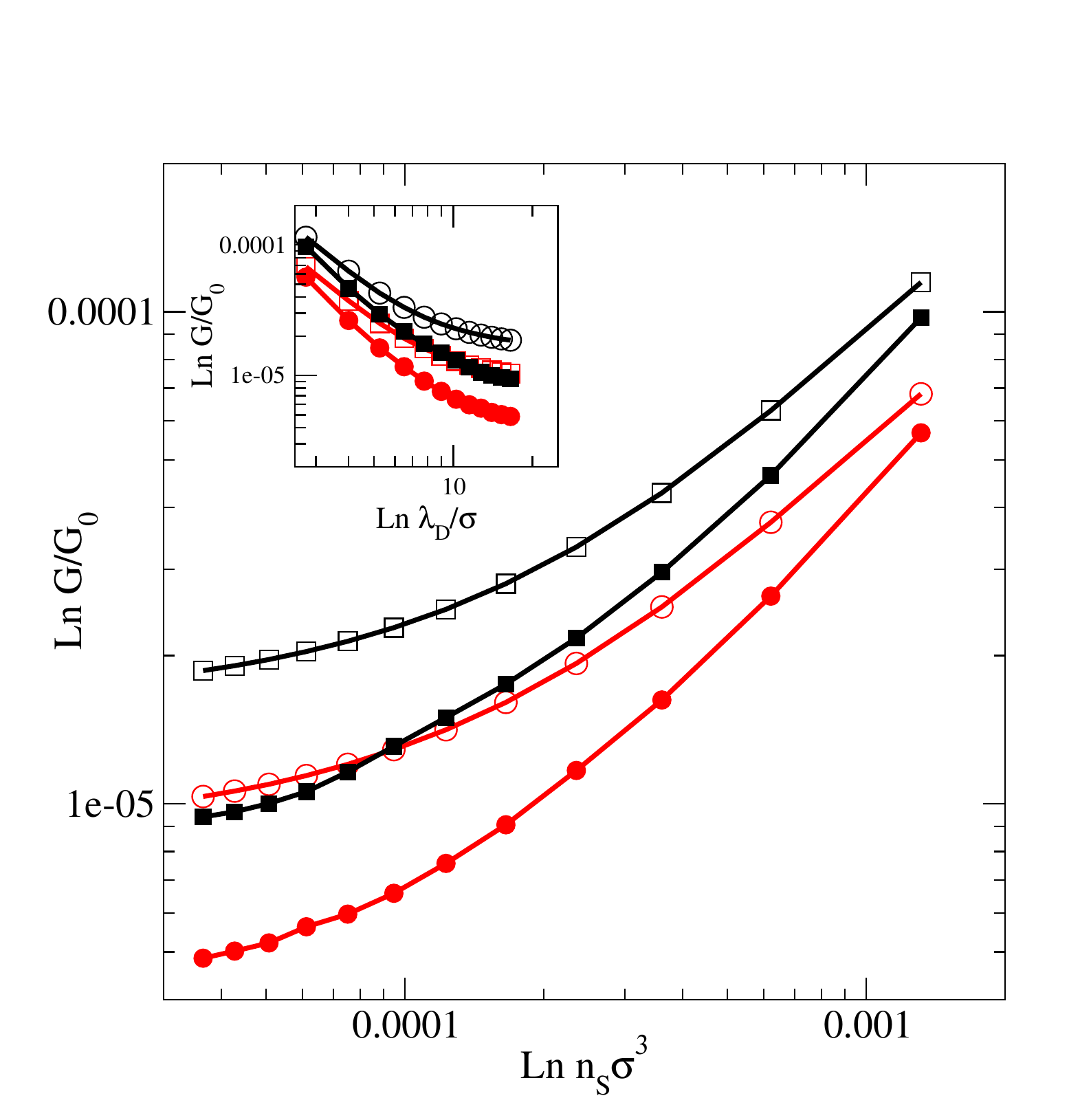}
\caption{Double logarithmic plot of the conductance, $G$, versus number density of carriers $n_s$ for  
model B (open symbols) and model C (filled symbols), 
in  the presence (circles) and in the absence (squares) of HS collisions.
The conductance is reported in units of $G_0 \equiv e^2 / v_T \sigma $.
The inset reports the conductance versus the Debye length.
}
\label{conductance}
\end{figure}

To demonstrate the bulk-like origin of electrical transport, 
in \ref{conductance} we report the numerical data for the electric conductance, $G$, 
for the geometries of models B and C. The simulations provide the ionic current 
$I$ and the potential difference $V$ applied at the ends of the channel, so we computed
the conductance according to the formula:
\be
I= G V.
\ee
An estimate of $G$ can be derived using the expression $G=\frac{\sigma_{el} S}{L}$,
where $S$ is the effective section of the channel, and given the geometry under consideration, $S/L=200/80=2.5$. 

\begin{figure}[htb]
\includegraphics[clip=true,width=12.0cm, keepaspectratio]{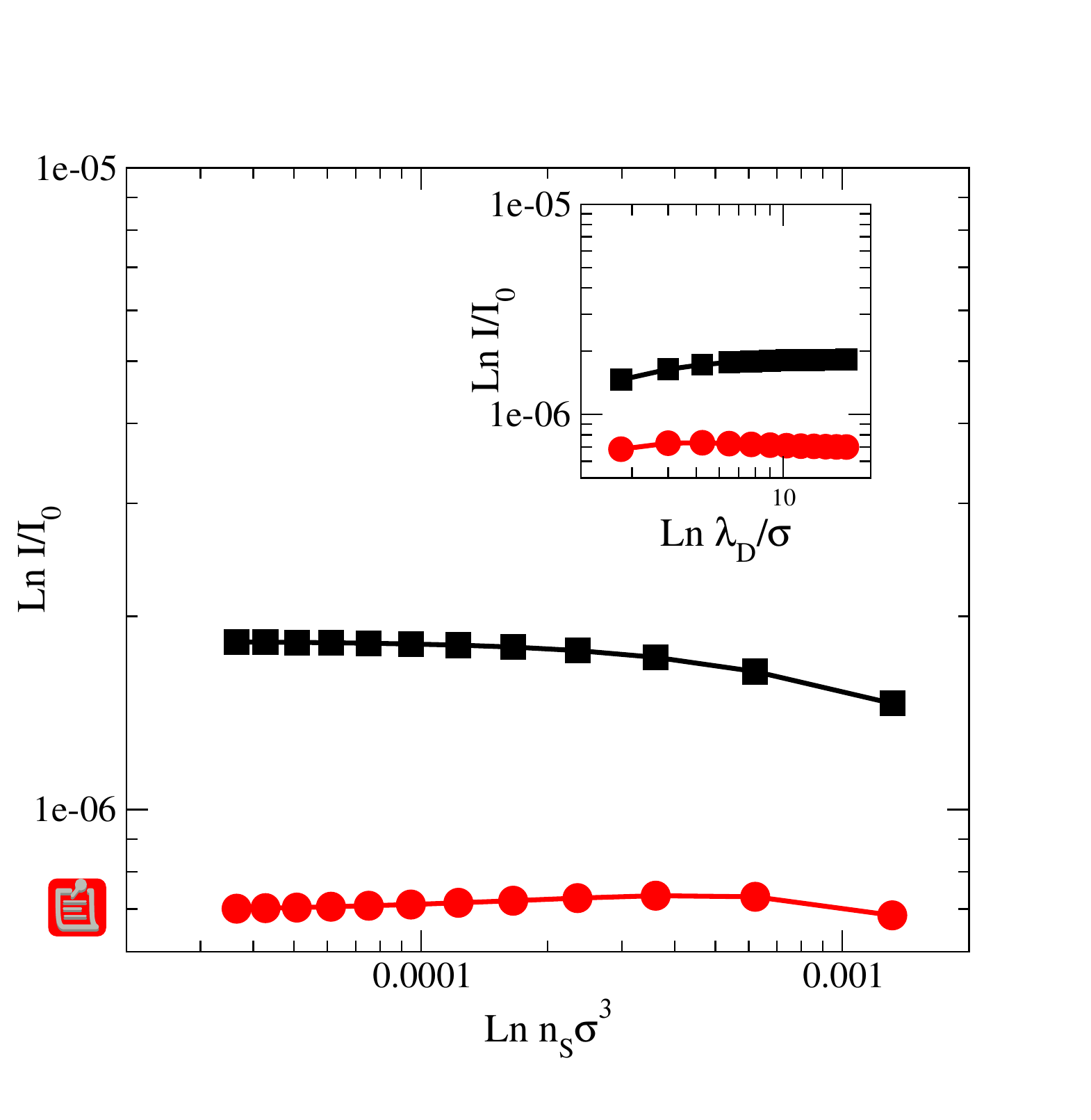}
\caption{Double logarithmic plot of the streaming current versus number density of carriers $n_s$ for  
model B in  the presence (circles) and in the absence (squares) of HS collisions.
The applied force is $8.33\times 10^{-7}$ in units of $v_T^2/\sigma$.
The ionic current is reported in units of $I_0 \equiv e v_T / \sigma $.
The inset reports the current versus the Debye length.
}
\label{streamingcurrent}
\end{figure}

\tr{\ref{streamingcurrent} reports the streaming current, that is, the ionic current
arising from a pressure gradient $\nabla p$ obtained for model B. 
The streaming current increases when increasing the Debye length, a behaviour
that is expected by considering that the ionic current is
proportional to the surface potential $\psi_s$, $I=\frac{\epsilon \psi_s S \nabla p}{\eta}$
\cite{Bocquet}, and that the surface potential is roughly proportional to the Debye length.
The presence of hard sphere collisions generates streaming currents smaller than without collisions,
due to the increased value of the dynamic viscosity. 
}

It should be kept in mind that, in order to compare the computed conductances 
\tr{arising from either a electric field or a pressure gradient} with 
the experimentally measured ones, the simulated system is the result of a periodic series of charged and
uncharged regions of the pore. Nevertheless, the simulation data are important as they show the intimate 
nature of conduction. As a matter of fact, as the bulk concentration decreases, the excess surface charges due to the presence
of the negatively charged surface of the channel becomes the leading contribution to the conduction mechanism. 
For very small values of $n_s$ the bulk formula
$\sigma_{el}= 2\frac{e^2}{\gamma }n_s$  (see  \ref{Drude}) does not hold and must be modified 
in order to account for the contribution due to the excess charge arising from the counterions
within the pore. By making the hypothesis that electroneutrality
is not a global constraint (see  \ref{englobal}) but rather holds locally, we simply replace
$n_s \to  n_s- \Sigma/e H$. The consequence of such replacement is the saturation of the conductance 
at low values of the electrolyte concentration $n_s$, as reproduced qualitatively in \ref{conductance}.
Finally, the operational value of $\lambda_D/\sigma =2.75$ presently studied corresponds to being relatively 
far from the saturation regime.


\section{Conclusions}
\label{Conclusions}

We have studied nanometric electro-osmotic flows by means of a self-consistent
kinetic description of the phase space distribution functions. 
The ternary mixture is composed by two oppositely charged species and a third neutral species 
representing the solvent, and is subject to external forces under non uniform conditions.
The short-range interaction between molecules has been modeled
by means of a hard sphere repulsion term and treated within the Revised Enskog theory. 
The Coulomb interaction has been treated within the mean field approximation, that is,
by disregarding charge induced correlations. 

The presented theory was shown to be consistent with the phenomenological equations of 
macroscopic electrokinetics, such as the Smoluchowski and Planck-Nernst-Poisson equations.
In essence, the approach  bridges the microscopic with the hydrodynamic level 
and allows to incorporate the inhomogeneity of the ionic solution determined by the presence 
of the confining surfaces.

{
In order to solve the system of equations for the multicomponent system, 
we have adopted a particular version of the 
Lattice Boltzmann method. We validated the theory and the numerical algorithm by considering 
the mass and charge transport under the effect of an applied electric fields in different channels having 
non uniform shapes and non uniform charging on the wall. 
}

{
In view of understanding the phenomenon of current rectification,
as observed in wedged channels having an orifice of nanometric size,
we have studied the departures from electroneutrality at molecular scale.
We have found that electroneutrality is restored along the longitudinal direction
within a Debye length. In addition, we observed mild undercharging of the wedged region at finite applied voltage.
The asymmetry of the curves for positive vs negative voltage is modest, ruling out the possibility
that this is at the origin of current rectification.
}

Besides the geometries presented here, the proposed methodology can be applied to a larger class 
of problems in electrofluidics.
For instance, it is possible to include semipermeable membranes, that is, 
channels impermeable to one type of ion and to the solvent. 
Other kinds of confining walls can be analyzed, such as chemically heterogeneous surfaces. 
A more comprehensive investigation of the role of differences in mass and size of the particles
is currently underway.

\section{Acknowledgments}
The authors are grateful to Z. Siwy, C. Holm, C. Pierleoni and J.-P. Hansen for illuminating discussions.




\begin{thebibliography}{99}



 \bibitem{denBerg} 
J.C.T. Eijkel, A van den Berg, 
{\em Microfluidics Nanofluidics}
 {\bf 1}, 249 (2005).

\bibitem{Sparreboom}
W. Sparreboom, A. van den Berg and J. C. T. Eijkel ,
{\em Principles and applications of nanofluidic transport},
Nature Nanotechnology {\bf 4}, 713, (2009).

\bibitem{Bocquet}
{\em Nanofluidics, from bulk to interfaces}
L.Bocquet and E.Charlaix, Chem. Soc. Rev. {\bf 39}, 1073, (2009).

\bibitem{Schoch}
R.B. Schoch, J.Han and P. Renaud,
{\em Transport phenomena in nanofluidics}
Rev.Mod.Phys. {\bf  80}, 839 (2008).

 \bibitem{Bruus}
H. Bruus, 
{\it Theoretical Microfluidics}, 
Oxford University Pres., New York, 2008 .

 \bibitem{Masliyah}
 J. H. Masliyah, S. Bhattacharjee
{\it Electrokinetic and colloid transport phenomena}
John Wiley and Sons, ( 2006).

    \bibitem{Kirby}
B. J. Kirby,  
{\it Micro and Nanoscale Fluid Mechanics}, 
Cambridge University Press (2010).


  \bibitem{Mukhopadhyay}
R. Mukhopadhyay, 
{\em Diving into droplets}
Anal. Chem. , {\bf 78}, 7379 (2006).



\bibitem{kontturi2008ionic}
{\it Ionic transport processes: in electrochemistry and membrane science},
 Kontturi, K. and Murtom{\"a}ki, L. and Manzanares, J.A., (2008) ,
Oxford University Press.

    
 \bibitem{Aluru}
R. Quiao and N.R. Aluru, 
{\em Ion concentrations and velocity profiles in nanochannel electroosmotic flows}
J. Chem. Phys. {\bf 118}, 4692 (2003).

\bibitem{Nilson}
R.H. Nilson and S.K. Griffiths, 
{\em Influence of atomistic physics on electro-osmotic flow: an analysis based on density functional theory}
J. Chem. Phys. {\bf 125}, 164510 (2006).

 \bibitem{Eisenberg}
B. Eisenberg, Y. Hyon and C. Liu,
{\em Energy variational analysis of ions in water and channels: Field theory for primitive models of complex ionic fluids}
 J. Chem. Phys. {\bf 133}, 104104 (2010).

 \bibitem{Singer}
 W. Zhu and S. J. Singer,  Z. Zheng and A. T. Conlisk, 
{\em Electro-osmotic flow of a model electrolyte}
Phys.  Rev. E, {\bf 71},  041501 (2005).


\bibitem{Press} 
W.H. Press et al., 
{\it Numerical Recipes in C: the art of scientific computing}, Cambridge University Press (2007).

 \bibitem{Karniadakis}
   G. Karniadakis, Ali Beskok, N.R. Aluru,  
{\it Microflows and nanoflows: fundamentals and simulation}
   Berlin  Springer, (2005).
   

  \bibitem{Tian}
F. Tian, B. Li, D. Y. Kwok, 
{\em Lattice Boltzmann Simulation of Electroosmotic flows in Micro- and Nanochannels}, 
Proceedings of the 2004 International Conference on MEMS, NANO and Smart Systems .

  \bibitem{Guo}
    Z. Guo, T. S. Zhao, Y. Shi, 
{\em A lattice Boltzmann algorithm for electro-osmotic flows in microfluidic devices}
J. Chem. Phys. 122, (2005 ).

    \bibitem{Wangwang}
{\em Lattice PoissonâBoltzmann simulations of electro-osmotic flows in microchannels}
 J. Wang, M. Wang, Z. Li, J. Coll. Interf. Sci. 296, (2006).

  \bibitem{Wangkang}
         M. Wang and  Q. Kang,   
{\em Modeling electrokinetic flows in microchannels using coupled lattice Boltzmann methods}
J. Computat. Phys. ,
{\bf 229},  728,     (2010).

\bibitem{Melchionnasucci}
S.~Melchionna and S. Succi, 
{\em  Electrorheology in nanopores via lattice Boltzmann simulation}
J. Chem. Phys. {\bf 120},  4492 (2004).

\bibitem{Melchionnaepl2011}
S.~Melchionna and U. Marini Bettolo Marconi,
{\em Electro-osmotic flows under nanoconfinement: A self-consistent approach}
Europhys.Lett {\bf 95}, 44002 (2011).


\bibitem{Vanbeijeren}
H. van Beijeren and M.H. Ernst, 
{\em The modified Enskog equation}
Physica A, {\bf 68}, 437 (1973).
 
\bibitem{UMBM2007}
U. Marini Bettolo Marconi and S.Melchionna,
{\em Phase-space approach to dynamical density functional theory}
J. Chem. Phys. {\bf  126}, 184109  (2007).
 
\bibitem{Simone2010}
U. Marini Bettolo Marconi and S.Melchionna,
{\em Dynamic density functional theory versus kinetic theory of simple fluids}
 J.Phys.:  Condens. Matter 
{\bf 22},  364110 (2010).



\bibitem{Rauscher2}
M. Rauscher, 
{\em DDFT for Brownian particles and hydrodynamics}
J.Phys.:  Condens. Matter  {\bf 22}, 364109 (2010).

 \bibitem{Brey}
J. W. Dufty, A. Santos, and J. Brey, 
{\em  Practical kinetic model for hard sphere dynamics}
Phys. Rev. Lett. ~{\bf 77}, 1270 (1996) .

 \bibitem{Santos}
 A.Santos, J.M. Montanero, J.W. Dufty and J.J. Brey, 
{\em Kinetic model for the hard-sphere fluid and solid}
Phys.Rev. E ~{\bf 57}, 1644 (1998).


\bibitem{LBgeneral}
S. Succi,
{\it The Lattice Boltzmann equation for fluid dynamics and beyond},  
1th edition , Oxford University Press, (2001).


\bibitem{BSV92}
R. Benzi, S. Succi, and M. Vergassola 
{\em The lattice Boltzmann equation: theory and applications}
Phys. Rep., {\bf 222} (1992) 145.


 \bibitem{Melchionna2008}
S.~Melchionna and U. Marini Bettolo Marconi,
{\em Lattice Boltzmann method for inhomogeneous fluids}
Europhys.Lett {\bf 81}, 34001 (2008).
 
\bibitem{Melchionna2009}
U. Marini Bettolo Marconi and S.Melchionna,
{\em Kinetic theory of correlated fluids: From dynamic density functional to Lattice Boltzmann methods}
J. Chem. Phys. {\bf 131},  014105 (2009).

\bibitem{Lausanne2010}
U. Marini Bettolo Marconi and S.Melchionna, 
{\em Multicomponent diffusion in nanosystems}
J. Chem. Phys  {\bf 135},  044104 (2011).

\bibitem{oleskyhansen}
The polarizability of the medium
could be accounted for by endowing the neutral component of the mixture  with a permanent dipole
as proposed by A.~Olesky and J.-P.~Hansen,
J. Chem. Phys., {\bf 132}, 204702 (2010).


 \bibitem{BGK}  
P. L. Bhatnagar, E. P. Gross, and M. Krook, 
{\em A model for collision processes in gases. I. Small amplitude processes in charged and neutral one-component systems}
Phys. Rev.  {\bf 94}, 511 (1954).

\bibitem{JCP2011}
U. Marini Bettolo Marconi and S.Melchionna,
{\em Dynamics of fluid mixtures in nanospaces}
 J. Chem. Phys {\bf 134}, 064118 (2011).

\bibitem{PRE2012}
S. Melchionna and U. Marini Bettolo Marconi
{\em Stabilized Lattice Boltzmann-Enskog method for compressible flows and its application to one and two-component fluids in nanochannels}
Phys. Rev. E, {\bf 85} 036707 (2012).

\bibitem{Fischer}
J. Fischer and M. Methfessel, 
{\em Born-Green-Yvon approach to the local densities of a fluid at interfaces}
Phys. Rev. A {\bf 22}, 2836 (1980).

 \bibitem{Boublik} 
 T. Boublik,  
{\em HardâSphere Equation of State}
J. Chem. Phys. {\bf  53}, 471 (1970). 

 \bibitem{Mansoori} 
 G.A. Mansoori, N.F. Carnahan, K.E. Starling and T.W. Leland,  
{\em Equilibrium thermodynamic properties of the mixture of hard spheres}
 J. Chem. Phys {\bf 54}, 1523 (1971).
 
   
  
  \bibitem{Joly} L. Joly, C. Ybert, E. Trizac and L. Bocquet,
{\em  Liquid friction on charged surfaces: From hydrodynamic slippage to electrokinetics}
  J. Chem.Phys. {\bf 125}, 204716, (2006). 
  
\bibitem{corry}
B. Corry, S. Kuyucak, S.-H. Chung, 
{\em Tests of continuum theories as models of ion channels. II. Poisson-Nernst-Planck theory versus Brownian dynamics}
Biophys. J. {\bf 78}, 2364 (2000).

 \bibitem{Kilic}
{\em Steric effects in the dynamics of electrolytes at large applied voltages. I. Double-layer charging}
 M.S. Kilic, M.Z. Bazant and A. Ajdari, Phys. Rev. E {\bf 75}, 021503 (2007).

\bibitem{frenkelsmit}
D. Frenkel, B Smit, 
{\it Understanding Molecular Simulation},
Academic Press, (2002).


\bibitem{Siwy1}
M.R. Powell, N. Sa, M. Davenport, K. Healy, I. Vlassiouk, S.E. Letant, L.A. Baker, Z.S. Siwy, 
{\em Noise Properties of Rectifying Nanopores}
J. Phys. Chem. C, {\bf 115}, 8775 (2011). 

\bibitem{Siwy2}
Z. Siwy and  A. Fulinski, 
{\em Fabrication of a synthetic nanopore ion pump}
Phys. Rev. Lett.  {\bf 89}, 198103, (2002).

\bibitem{Cervera}
J. Cervera, B. Schiedt and P. Ramirez,
{\em A Poisson/Nernst-Planck model for ionic transport through synthetic conical nanopores}
Europhys. Lett., {\bf 71}, 35, (2005).

\bibitem{Dietrich}
P. Bryk, R. Roth, M. Schoen and S. Dietrich,
{\em Depletion potentials near geometrically structured substrates}
Europhys. Lett. \textbf{ 63} 233 (2003).

\bibitem{WOERMANN}
D. Woermann, 
{\em Electrochemical transport properties of a cone-shaped nanopore: revisited}
Phys. Chem. Chem. Phys. \textbf{6} 3130 (2004). 

\end{thebibliography}
 \end{document}